\documentclass[12pt]{iopart}
\usepackage{amssymb}
\usepackage{epsfig}
\newcommand{\inp}[3]{\left\langle #1\left|#2\right|#3\right\rangle}
\newcommand{\ket}[1]{\left|#1\right\rangle} 
\newcommand{\bra}[1]{\left\langle#1\right|} 
\newcommand{\braket}[2]{\left\langle#1|#2\right\rangle}

\newcommand{\comm}[2]{\left[#1,#2\right]}
\newcommand{\acomm}[2]{\left\{#1,#2\right\}}
\newcommand{\ave}[1]{\left<#1\right>}
\newcommand{\afg}[2]{\frac{d#1}{d#2}}
\newcommand{\pafg}[2]{\frac{\partial #1}{\partial #2}}

\begin{document}

\title[Non-perturbative flow equations]{Non-perturbative flow equations from continuous unitary transformations}
\author{J N Kriel, A Y Morozov and F G Scholtz}
\address{Institute of Theoretical Physics, University of Stellenbosch, South Africa}

\begin{abstract}

We use a novel parameterization of the flowing Hamiltonian to show that the flow equations based on continuous unitary transformations, as proposed by Wegner, can be implemented through a nonlinear partial differential equation involving one flow parameter and two system specific auxiliary variables.  The implementation is non-perturbative as the partial differential equation involves a systematic expansion in fluctuations, controlled by the size of the system, rather than the coupling constant.  The method is applied to the Lipkin model to construct a mapping which maps the non-interacting spectrum onto the interacting spectrum to a very high accuracy.  This function is universal in the sense that the full spectrum for any (large) number of particles can be obtained from it. 
In a similar way expectation values for a large class of operators can be obtained, which also makes it possible to probe the stucture of the eigenstates.
  
\end{abstract}


\maketitle

\section{Introduction and background}
\label{Intro}
The renormalization group and associated flow equations \cite{Zinn} is one of the cornerstones of modern physics and one of the very few potentially non-perturbative techniques available to us.  Not only does the renormalization group allow the construction of low-energy effective theories for interacting quantum systems, but it also allows us to probe the phase structure of such systems. It is probably for this reason that considerable interest was generated in the flow equations recently proposed by Wegner \cite{Wegner} and independently by Glazek and Wilson \cite{Glazeck} as well as its demonstrated close relation to the renormalization group \cite{Glazeck}. Further applications of the flow equations, based on successive infinitesimal unitary transformations, to diverse quantum mechanical problems followed, which included the treatment of the electron-phonon coupling \cite{Wegner}, the spin-boson Hamiltonian \cite{Kehrein}, the Hubbard model \cite{Stein1}, the Sine-Gordon model \cite{Kehrein1} and the Foldy-Wouthuysen transformation \cite{Bylev}.  The Lipkin model has been particularly prominent among applications \cite{PirnerFriman,Stein2, Mielke}, probably because it is a many-body model that can be solved numerically quite easily, but which exhibits a non-trivial quantum phase transition \cite{Lipkin} and could therefore shed light on the practical implementation of these flow equations in a simple, and yet non-trivial setting.

One difficulty encountered in the practical implementation of the renormalization group and flow equations is that the Hamiltonian does not preserve its form under the flow and that additional operators, not present in the original Hamiltonian, are generated \cite{Zinn,Wegner}.  The generic situation therefore yields an infinite set of nonlinear coupled differential equations for the coupling constants, and some form of truncation is required.  One way of doing this is through a perturbative implementation of the renormalization group equations, which corresponds to a systematic expansion in the coupling constant. Of course this procedure becomes invalid when flow to strong coupling occurs and non-perturbative aspects of the problem at hand cannot be probed in this way.  In the context of the Wegner flow equation this issue has been dealt with in all the applications mentioned above by parameterizing the flowing Hamiltonian and then attempting to close on a finite set of equations through some approximation.  This procedure failed completely in situations where the system exhibited non-perturbative features, such as the quantum phase transition in the Lipkin model \cite{PirnerFriman}, where the coupling simply flows to infinity.  A more detailed discussion on the role of truncation in the accuracy of the flow equation can be found in \cite{Stauber}.    

In a previous paper one of the present authors suggested a self-consistent implementation of the Wegner flow equation which yielded good results in both phases of the Lipkin model \cite{FGS}.  The central idea of this approach was a systematic expansion of the flow equation in the fluctuations around the ground state expectation value, which is controlled by the inverse of the number of particles in the system, instead of the coupling constant.  This  resulted in an effective non-perturbative implementation of the flow equation, explaining its success in describing both phases of the Lipkin model.  However, the approach as outlined in \cite{FGS} made use of specific features of the Lipkin model and was only successful in describing the low-energy behaviour, while it fails in producing the excitation energies of higher states correctly.  Here we describe a generalization of this approach which can be applied to any finite dimensional Hamiltonian (see the next section for further elaboration) and which succeeds in producing the full spectrum and expectation values correctly. 
    
To set the scene and the notation we start with a brief review of the Wegner flow equation.  The central idea behind the flow equation is to diagonalize the Hamiltonian with a family of successive infinitesimal unitary transformations, $U(\ell)$, parameterized by a flow parameter $\ell$ and generated by some appropriate anti-hermitian operator $\eta(\ell)$, i.e., $\frac{dU(\ell)}{d\ell}=-U(\ell)\eta(\ell)$.  The resulting equation that determines the flow of the transformed Hamiltonian $H(\ell)= U^\dagger(\ell)H(0)U(\ell)$ is
\begin{equation}
\label{flow}
\frac{d H(\ell)}{d\ell}=[\eta(\ell),H(\ell)]\,.
\end{equation} 
 
In Wegner's original formulation the generator $\eta(\ell)$ was chosen as the commutator of the diagonal part of the flowing Hamiltonian $H(\ell)$ in some basis with the Hamiltonian itself, i.e., $\eta(\ell)=[{\rm diag}(H(\ell)),H(\ell)]$ and it was shown that the final Hamiltonian $H(\infty)$ will be diagonal in the chosen basis \cite{Wegner}.  Here we allow a slightly more general formulation of the flow equation and consider a generator $\eta(\ell)=[H_0,H(\ell)]$ where it is assumed that the eigenvalues and eigenstates of $H_0$ are known (note that $H_0$ is $\ell$ independent). It is easy to prove that the final Hamiltonian $H(\infty)$ is diagonal in the basis of $H_0$ if the spectrum of $H_0$ is non-degenerate, else it will assume a block diagonal structure.  The proof rests on the following observation:
\begin{equation} \label{diagproof}
\frac{d}{d\ell}{\rm tr}(H(\ell)-H_0)^2=-2{\rm tr}([H_0,H(\ell)]^\dagger[H_0,H(\ell)])<0. 
\end{equation}
Thus ${\rm tr}(H(\ell)-H_0)^2$ is a monotonically decreasing function of $\ell$ which is bounded from below by zero.  This implies that the derivative must vanish in the $\ell\rightarrow\infty$ limit, and since the right-hand side of (\ref{diagproof}) is simply the trace norm of $[H_0,H(\ell)]$ this in turn implies that $[H_0,H(\infty)]=0$.  Furthermore it can be shown \cite{Brockett} that the eigenvalues of $H=H(0)$, as they  appear on the diagonal of $H(\infty)$, have the same ordering as the eigenvalues of $H_0$.  

In section 2 we describe the general parameterization on which our approach is based and outline the general derivation of the flow equation as an expansion in terms of the fluctuations. We briefly discuss the general properties of the resulting flow equation, which now takes the form of an ordinary (in contrast to operator valued) nonlinear partial differential equation, and describe how the spectrum and expectation values can be computed. In section 3 we apply this method to the Lipkin model where the calculations can be done simply and explicitly. In section 4 the results obtained by solving the resulting flow equation numerically are compared in detail with the exact results. The flow equation provides some interesting new perspectives on the phase transition and the general properties of the spectrum in the second phase. We conclude in section 5 with a summary, an outline of future applications and outstanding issues. The technicalities and proofs are given in three appendices.

\section{A general parametrization of flowing operators}
\label{genrepsec}
We consider the flow of a real, hermitian operator $H(\ell)$ which acts on a finite dimensional Hilbert space ${\cal H}$. Here we have in mind applications to systems with a finite (but large) number of degrees of freedom such as interacting fermions or spins on a finite sized lattice where one is in particular interested in studying the behaviour of the system as a function of size.  For the moment we only consider systems in which there is no breaking of time reversal symmetry.  Breaking of time reversal symmetry can be dealt with at the expense of a generalization of the expansion (\ref{fexpandflux}) and correspondingly the relations (\ref{2ndorder}) discussed below. 

The restriction to a finite dimensional Hilbert space is made here since the expansion (\ref{expansion}) below is easily proven in this case (see \ref{appa} ).  It is, however, by no means essential.  If the Hilbert space is infinite dimensional the expansion (\ref{expansion}) still applies to bounded operators \cite{Blank} and one may study the flow of a bounded function of the Hamiltonian, instead of the Hamiltonian itself. Alternatively one can introduce a cut-off in some basis, e.g., a momentum cut-off and study the behaviour of the system as a function of the cut-off. 

As mentioned in the introduction the usual method of closing the flow equation consists of replacing $H(\ell)$ by an approximate parameterized form, and then solving for the parameters as functions of $\ell$. This approach invariably fails at large coupling or when the system exhibits non-perturbative features as the approximate parameterization of $H(\ell)$ breaks down. We abandon this approach in favour of a representation for $H(\ell)$ which is both exact and very general.
Let $H_0$ and $H_1$ be hermitian operators acting on ${\cal H}$ which together form an irreducible set.  What follows holds for any irreducible set, however, the case of two operators appear naturally in flow equations as the Hamiltonian can be written in the form $H=H_0+H_1$ where the spectrum and eigenstates of $H_0$ are known.  Note that if $H_0$ and $H_1$ are reducible on ${\cal H}$, but irreducible on a proper subspace of ${\cal H}$, the problem can be restricted to this smaller subspace making the irreducibility of $H_0$ and $H_1$ a very natural requirement. In \ref{appa} it is shown that any operator acting on ${\cal H}$ can be written as a polynomial in $H_0$ and $H_1$. In particular this holds for $H(\ell)$ so we may write
\begin{equation}
\label{expansion}
	H(\ell)=\Gamma+\Gamma^iH_i+\Gamma^{ij}H_iH_j+\Gamma^{ijk}H_iH_jH_k+\ldots,
	\label{genparamh}
\end{equation}
where each $\Gamma^{ij\ldots}$ coefficient is a function of $\ell$ and repeated indices indicate sums over $0$ and $1$. Since $H(\ell)$ is both hermitian and real it follows that the $\Gamma$'s are invariant under the reversal of indices:  
\begin{equation}
	\Gamma^{n_1n_2\ldots n_k}=\Gamma^{n_kn_{k-1}\ldots n_1}\ \ \ \forall k>0,\ n_i\in\{0,1\}.
	\label{hermiteconstr}
\end{equation}
Now define $\Delta H_i=H_i-\ave{H_i}$ as the fluctuation of $H_i$ around the expectation value $\ave{H_i}=\inp{\phi}{H_i}{\phi}$, where $\ket{\phi}$ is some arbitrary state. By setting $H_i=\Delta H_i+\ave{H_i}$ in equation (\ref{genparamh}) we obtain $H(\ell)$ as an expansion in these fluctuations 
\begin{equation}
\label{expansion1}
	H(\ell)=\bar{\Gamma}+\bar{\Gamma}^i\Delta H_i+\bar{\Gamma}^{ij}\Delta H_i\Delta H_j+\bar{\Gamma}^{ijk}\Delta H_i\Delta H_j\Delta H_k+\ldots,
\end{equation}
where
\begin{eqnarray}
\bar{\Gamma}=\Gamma+\Gamma^i\ave{H_i}+\Gamma^{ij}\ave{H_i}\ave{H_j}+\Gamma^{ijk}\ave{H_i}\ave{H_j}\ave{H_k}+\ldots \label{gammabardef}\\
\bar{\Gamma}^i=\Gamma^i+\left(\Gamma^{ij}+\Gamma^{ji}\right)\ave{H_j}+\left(\Gamma^{ijk}+\Gamma^{jik}+\Gamma^{kji}\right)\ave{H_j}\ave{H_k}+\ldots\\
\bar{\Gamma}^{ij}=\Gamma^{ij}+\left(\Gamma^{kij}+\Gamma^{ikj}+\Gamma^{ijk}\right)\ave{H_k}+\ldots
\end{eqnarray}
and so forth. 

Note that $\bar{\Gamma}$, when viewed as a function of $\ell$, $\ave{H_0}$ and $\ave{H_1}$, encodes information about all the expansion coefficients $\Gamma^{ij\ldots}$ appearing in equation (\ref{expansion}). A natural strategy that presents itself is to set up an equation for $\bar{\Gamma}$ as a function of $\ell$, $\ave{H_0}$ and $\ave{H_1}$ in some domain ${\cal D}$ of the $\ave{H_0}$--$\ave{H_1}$ plane, determined by the properties of the operators $H_0$ and $H_1$.  In particular we require that this domain includes values of $\ave{H_0}$ ranging from the largest to the smallest eigenvalues of $H_0$.  For this purpose we require a family of states $\ket{\phi(\lambda)}$ parameterized by a continuous set of variables $\lambda\in \bar {\cal D}$ such that as $\lambda$ is varied over the domain $\bar {\cal D}$, $\ave{H_0}$ and $\ave{H_1}$ range continuously over the domain ${\cal D}$. A very general set of states that meets these requirements are coherent states \cite{Klauder}. In this representation we may consider $\ave{H_0}$ and $\ave{H_1}$ as continuous variables with each $\bar{\Gamma}$ a function of $\ave{H_0}$, $\ave{H_1}$ and the following relations hold generally: 
\begin{equation}
\label{2ndorder}
	\bar{\Gamma}^i=\frac{\partial\bar{\Gamma}}{\partial \ave{H_i}}\ \ {\rm and}\ \ \bar{\Gamma}^{ij}=\frac{1}{2}\frac{\partial^2 \bar{\Gamma}}{\partial\ave{H_i}\partial\ave{H_j}}.
\end{equation}
Coefficients with three or more indices cannot be written in this way, since, for example, $\bar{\Gamma}^{101}$ need not be equal to $\bar{\Gamma}^{011}$. The general relationship between the true coefficients and derivatives of $\bar{\Gamma}$ are
\begin{equation}
 \frac{\partial^{n+m} \bar{\Gamma}}{\partial\ave{H_0}^n\partial\ave{H_1}^m}=n!m!\sum_{p}\bar{\Gamma}^{p(n,m)}
\end{equation}
where the sum over $p$ is over all the distinct ways of ordering $n$ zeros and $m$ ones. For example
\begin{equation}
	\frac{1}{3!}\frac{\partial^{3} \bar{\Gamma}}{\partial\ave{H_0}^2\partial\ave{H_1}^1}=\frac{1}{3}\left(\bar{\Gamma}^{001}+\bar{\Gamma}^{010}+\bar{\Gamma}^{100}\right).
\end{equation}
Replacing $\bar{\Gamma}^{001}$ by the left-hand side of the equation above is equivalent to approximating $\bar{\Gamma}^{001}$ by the average of $\bar{\Gamma}^{001}$,$\bar{\Gamma}^{010}$ and $\bar{\Gamma}^{100}$. In general this amounts to approximating $\bar{\Gamma}^{n_1\ldots n_\ell}$ by the average of the coefficients corresponding to all the distinct reorderings of $n_1,\ldots,n_\ell$. 

To set up an equation for $\bar{\Gamma}(\ave{H_0},\ave{H_1},\ell)$ we insert the expansion (\ref{expansion1}) into the flow equation (\ref{flow}) with $\eta(\ell)=[H_0,H(\ell)]$ and take the expectation value with respect to the state $\ket{\phi(\lambda)}$. The left-and right-hand sides become a systematic expansion in orders of the fluctuations $\ave{\Delta H_{n_1}\ldots \Delta H_{n_\ell}}$:
\pagebreak
\begin{eqnarray}
\label{flow1}
\frac{\partial\bar{\Gamma}}{\partial\ell}+\frac{\partial\bar{\Gamma}^{ij}}{\partial\ell}\ave{\Delta H_i\Delta H_j}+\ldots&=&\bar{\Gamma}^{i}\bar{\Gamma}^{j}\ave{[[\Delta H_0,\Delta H_{i}],\Delta H_{j}]} \nonumber\\ 
&+&\bar{\Gamma}^{i}\bar{\Gamma}^{j_1 j_2}\ave{[[\Delta H_0,\Delta H_{i}],\Delta H_{j_1}\Delta H_{j_2}]}\nonumber\\
&+&\bar{\Gamma}^{i_1 i_2}\bar{\Gamma}^{j}\ave{[[\Delta H_0,\Delta H_{i_1}\Delta H_{i_2}],\Delta H_{j}]}\\ 
&+&\bar{\Gamma}^{i_1 i_2}\bar{\Gamma}^{j_1 j_2}\ave{[[\Delta H_0,\Delta H_{i_1}\Delta H_{i_2}],\Delta H_{j_1}\Delta H_{j_2}]}\nonumber\\ 
&+&\ldots\nonumber
\end{eqnarray}
Note that writing $H_0$ or $\Delta H_0$ in the first position of the double commutator on the right-hand side of (\ref{flow1}) is a matter of taste. The expectation values appearing above are naturally functions of $\lambda$ and may be written as functions of $\ave{H_0}$ and $\ave{H_{1}}$ by inverting the equations $\ave{H_i}=\inp{\phi(\lambda)}{H_i}{\phi(\lambda)}$ to obtain $\lambda(\ave{H_0},\ave{H_1})$. With an appropriately chosen state, such as a coherent state, the higher orders in the fluctuation can often be neglected, as the expansion is controlled by the inverse of the number of degrees of freedom (see \ref{appc} for an explicit example).  Upon doing this we note that replacing the ${\bar\Gamma^{ij\ldots}}$'s with more than two indices by a derivative will introduce corrections in (\ref{flow1}) of an order higher than the terms already listed, or, on the level of the operator expansion, corrections higher than second order in the fluctuations. Working to second order in the fluctuations we can therefore safely replace ${\bar\Gamma^{ij\ldots}}$ by the derivatives of a function $f\left(\ave{H_0},\ave{H_1},\ell\right)\equiv \bar{\Gamma}(\ave{H_0},\ave{H_1},\ell)$ and write for $H(\ell)$:
\begin{eqnarray}
	\fl H(\ell)&=&f+f^{10}\Delta H_0+f^{01}\Delta H_1+\frac{1}{2}f^{11}\left\{\Delta H_0,\Delta H_1\right\}+\frac{1}{2}f^{20}\Delta H_0^2+\frac{1}{2}f^{02}\Delta H_1^2+\cdots \nonumber \\
	\fl &=&\sum_{i,j=0}^\infty\frac{f^{ij}}{(i+j)!}\sum\left({\rm Distinct\ orderings\ of\ } i\ \Delta H_0{\rm 's}\ {\rm and}\ j\ \Delta H_1's \right)
	\label{fexpandflux}
\end{eqnarray}
where $\{\cdot,\cdot\}$ denotes the anti-commutator and
\begin{equation}
	 f^{ij}=\frac{\partial^{\,i+j} f}{\partial\ave{H_0}^i\partial\ave{H_1}^j}.
\end{equation}
This turns the flow equation (\ref{flow1}) into a nonlinear partial differential equation for $f\left(\ave{H_0},\ave{H_1},\ell\right)$, correct up to the order shown in (\ref{flow1}).  The choice of coherent state, the corresponding calculation of the fluctuations appearing in equation (\ref{flow1}) and the identification of the parameter controlling the expansion are problem specific.  Therefore we focus for the rest of this paper on a specific example, namely the Lipkin model, where all the calculations can be done explicitly.  We would, however, like to emphasize that the results obtained here for the Lipkin model go well beyond any results previously obtained for the Lipkin or any other model within the context of the Wegner flow equations. 

Before embarking on the discussion of the Lipkin model, there are a number of general statements that can be made about the flow equation (\ref{flow1}) and the behaviour of $f\left(\ave{H_0},\ave{H_1},\ell\right)$.  The first property to be noted is that since $H(\infty)$ is diagonal in the eigenbasis of $H_0$ it should become a function of only $H_0$, provided that the spectrum of $H_0$ is non-degenerate. This is reflected in the behaviour of $f$ by the fact that $f\left(\ave{H_0},\ave{H_1},\ell=\infty\right)$ should be a function of $\ave{H_0}$ only.  This is indeed borne out to high accuracy in our later numerical investigations. This function in turn provides us with the functional dependence of $H(\infty)$ on $H_0$. Keeping in mind the unitary connection between $H(\infty)$ and $H$ this enables us to compute the eigenvalues of $H$ straightforwardly by inserting the supposedly known eigenvalues of $H_0$ into the function $f\left(\ave{H_0},\infty\right)$.    

A second point to note is that the considerations above apply to the flow of an arbitrary real hermitian operator.  It is easily verified that the transformed operator $P(\ell)= U^\dagger(\ell)P(0)U(\ell)$ satisfies the flow equation:
\begin{equation}
\label{flow2}
\frac{d P(\ell)}{d\ell}=[\eta(\ell),P(\ell)]\,
\end{equation} 
where the same choice of $\eta(\ell)$ as for the Hamiltonian has to be made, i.e., $\eta(\ell)=[H_0,H(\ell)]$.  The expansion (\ref{fexpandflux}) can be made for both operators $H(\ell)$ and $P(\ell)$.  Denoting the corresponding function for $P(\ell)$ by $g\left(\ave{H_0},\ave{H_1},\ell\right)$ the flow equation (\ref{flow2}) turns into a linear partial differential equation for $g\left(\ave{H_0},\ave{H_1},\ell\right)$ containing the function $f\left(\ave{H_0},\ave{H_1},\ell\right)$, which is determined by the nonlinear partial differential equation (\ref{flow1}). The expectation value of $P=P(0)$ in an eigenstate $\ket{E_n}$ of $H$ can be expressed as
\begin{equation}
	\inp{E_n}{P}{E_n}=\inp{E_n}{U(\ell)U^\dag(\ell) PU(\ell)U^\dag(\ell)}{E_n}=\inp{E_n,\ell}{P(\ell)}{E_n,\ell}
\end{equation}
where $\ket{E_n,\ell}=U^\dag(\ell)\ket{E_n}$. In the limit $\ell\rightarrow\infty$ the states  $\ket{E_n,\infty}$ are simply the eigenstates of $H_0$, which are supposedly known.  In this way the computation of the expectation value $\inp{E_n}{P}{E_n}$ can be translated into the calculation of expectation values of the operator $P(\infty)$, obtained by solving the flow equation (\ref{flow2}), in the known eigenstates of $H_0$.   

\section{The Lipkin Model}
\label{lipkinmodel}
The Lipkin model describes $N$ fermions distributed over two $\Omega$-fold degenerate levels separated by an energy of $\xi$. The interaction $V$ introduces scattering of pairs between shells. Labeling the two levels by $\sigma=\pm1$, the Hamiltonian reads
\begin{equation}
 H=\frac{\xi}{2} \sum_{\sigma,p}\sigma a^\dag_{p,\sigma} a_{p,\sigma}+\frac{V}{2}\sum_{p,p',\sigma}a^\dag_{p,\sigma}a^\dag_{p',\sigma}a_{p',-\sigma}a_{p,-\sigma}
\end{equation}
where the sum over $p,p^\prime$ runs over the level degeneracy $1\ldots\Omega$. A spin representation for $H$ can be found by introducing the ${\rm su}(2)$ generators
\begin{equation}
J_z=\frac{1}{2}\sum_{\sigma,p}\sigma a^\dag_{p,\sigma} a_{p,\sigma} \ \ \ {\rm and} \ \ \ J_\pm=\sum_{p}\sigma a^\dag_{p,\pm \sigma} a_{p,\mp \sigma}
\end{equation}
which satisfy $\left[J_z,J_\pm\right]=\pm J_{\pm}$ and $\left[J_+,J_-\right]=2J_{z}$.
We divide $H$ by $\xi$ and define the dimensionless coupling constant $\beta=2jV/\xi$ to obtain
\begin{equation}
	H=J_z+\frac{\beta}{4 j}\left(J_+^2+J_-^2\right),
\end{equation}
where all energies are now expressed in units of $\xi$. 
Since $\left[H,J^2\right]=0$ the Hamiltonian acts within irreducible representations of ${\rm su}(2)$, leading to a block diagonal structure of size $2j+1$. The low-lying states occur in the multiplet $j=N/2$. When $\beta=0$ the ground state is simply $\ket{G}=a^\dag_{1,-1}\ldots a^\dag_{N,-1}\ket{0}$ which is written as $\ket{G}=\ket{j=N/2,-j}$ in the spin basis. Non-zero values of $\beta$ cause particle-hole excitations across the gap, and at $\beta=\pm 1$ the model shows a phase transition from an undeformed first phase to a deformed second phase. To distinguish the two phases we use the order parameter $\Omega\equiv1+\left<J_z\right>/j$ where $\left<J_z\right>$ is the expectation value of $J_z$ in the ground state. A discussion of this model and its features appear in \cite{Lipkin}. We use the shorthand $\ket{m}\equiv\ket{j,m}$ for the eigenstates of $J_z$ and denote the eigenstate of $H$ with energy $E_n$ by $\ket{E_n}$ where $E_0\leq E_1\leq E_2\ldots$.
\subsection{Flow equations for the Lipkin Model}
\label{lipkinflow}
Setting
\begin{eqnarray}
	H_0=J_z\ \ {\rm and}\ \ H_1=\frac{1}{4j}\left(J_+^2+J_-^2\right)
\end{eqnarray}
the Hamiltonian becomes $H=H_0+\beta H_1$.
We will consider flow equations of the form
\begin{equation}
	\frac{d H(\ell)}{d \ell}=\comm{\eta(\ell)}{H(\ell)},\ \ \ H(0)=H_0+\beta H_1
	\label{floweqdef}
\end{equation}
where the generator is chosen as in \cite{PirnerFriman},
\begin{equation}
	\eta(\ell)=\comm{J_z}{H(\ell)}=\comm{H_0}{H(\ell)}.
	\label{geneqdef}
\end{equation}
The choice of $J_z$ in the generator has two important consequences:
\begin{enumerate}
	\item Since $J_z$ is non-degenerate and $\eta(\infty)=0$, $H(\infty)$ must be diagonal in the $J_z$ basis.
	\item $H(\ell)$ will remain band-diagonal, i.e. $\inp{n}{H(\ell)}{m}=0$ when $\left|n-m\right|\notin\{0,2\}$. 
\end{enumerate}
The unitary transformation $U(\ell)$ defined by
\begin{equation}
	\afg{U(\ell)}{\ell}=-U(\ell)\eta(\ell)\ \ {\rm and}\ \ U(0)=I
	\label{DEU}
\end{equation}
relates $H(\ell)$ to $H(0)$ in the usual way:
\begin{equation}
	H(\ell)=U^\dag(\ell)H(0)U(\ell).
\end{equation}
The eigenvalues of $H$ appear on the diagonal of $H(\infty)$ in the same order as in $J_z$ \cite{Brockett}, i.e. increasing from top to bottom. It follows that $E_n=\inp{-j+n}{H(\infty)}{-j+n}$ for each $n=0,\ldots,2j$. The eigenstates of $H(\ell)$ evolve as $\ket{E_n,\ell}=U^\dag(\ell)\ket{E_n}$ during the flow, and since $E_n=\inp{E_n}{H}{E_n}=\inp{E_n,\infty}{H(\infty)}{E_n,\infty}$ we conclude that $\ket{E_n,\infty}=\ket{-j+n}$.
\subsection{Representing the flowing operators}
\label{flowrep}
As suggested by the notation, we choose to represent flowing operators in terms of $H_0$ and $H_1$ as described in section \ref{genrepsec}. However $\{H_0,H_1\}$ is not an irreducible set, since both leave the subspace spanned by $\left\{\ket{n}\,|\,n\ {\rm even}\right\}$ invariant. As pointed out earlier this is easily remedied by restricting ourselves to this subspace.  However, it turns out that in this case an expansion of the form (\ref{expansion}) can, as a consequence of the particular choice of the generator, even be found in the reducible case.  Indeed, we may write $H(\ell)$ as
\begin{equation}
\label{spespar}
	H(\ell)=\sum_{i=0}d_i(\ell)H^i_0+\sum_{i=0} n_i(\ell)H^i_0 H_1 H^i_0.
	\label{specialparamh}
\end{equation}
where the $d_i$'s and $n_i$'s are the diagonal and off-diagonal coefficients respectively. This can be verified by inspection of the flow equation which shows that no new terms are generated. We will not use this model dependent parameterization in what follows, and only present it as proof that such a representation exists, demonstrating the more general theorem in this particular model. This result extends to $U(\ell)$ by equation (\ref{DEU}), and then to any flowing operator $P(\ell)=U^\dag(\ell)P(0)U(\ell)$, provided that $P(0)$ can be expressed in terms of $H_0$ and $H_1$. 

\begin{figure}[t]
\begin{center}
	\epsfig{file=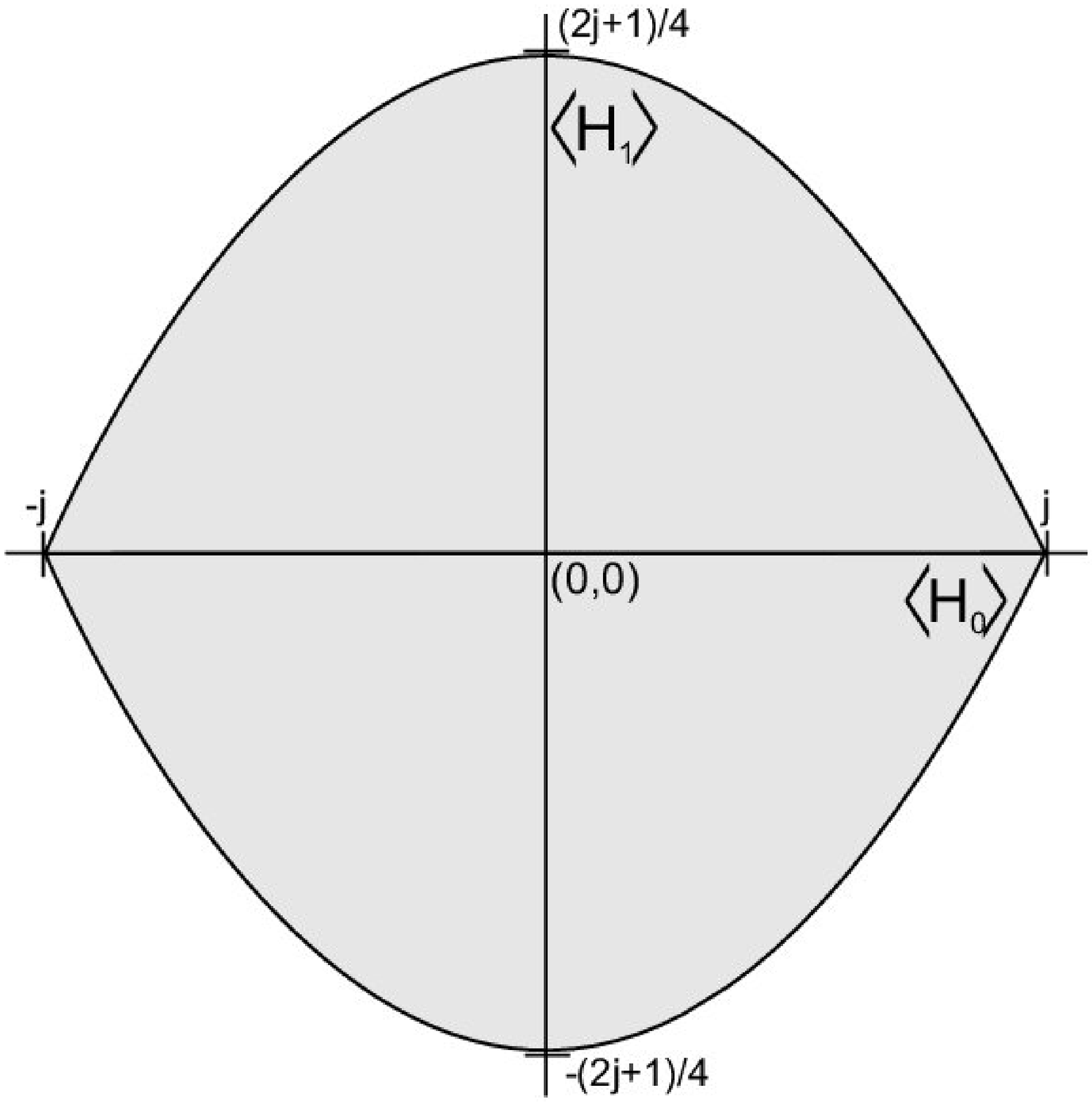,height=6.0cm,clip=,angle=0}
	\caption{The domain of $f(\ave{H_0},\ave{H_1},\ell)$.}
	\label{domainfig}
\end{center}
\end{figure}

We will calculate the averages of $H_0$ and $H_1$ with respect to the coherent state
\begin{eqnarray}
\label{cohstate}
\ket{z}\equiv N^{-j}\exp(z J_+)\ket{-j},
\end{eqnarray}
where $z=r\exp(i\theta)\in\mathbb{C}$ and $N=1+r^2$. Using the method described in \ref{appb} it is found that
\begin{equation}
	\ave{H_0}=j\left(\frac{r^2-1}{r^2+1}\right)\ \ {\rm and}\ \ \ave{H_1}=\frac{(2j-1)\,r^2\cos(2\theta)}{(1+r^2)^2}.
\end{equation}
The possible values of $\left(\ave{H_0},\ave{H_1}\right)$ are shown in figure \ref{domainfig}. 
With these definitions in place we conclude that the flowing Lipkin Hamiltonian may be written as
\begin{eqnarray}
	\fl H(\ell)=f+f^{10}\Delta H_0+f^{01}\Delta H_1+\frac{1}{2}f^{11}\left\{\Delta H_0,\Delta H_1\right\}+\frac{1}{2}f^{20}\Delta H_0^2+\frac{1}{2}f^{02}\Delta H_1^2+\cdots
	\label{lipkinhexp}
	\label{kaas}
\end{eqnarray}
with $H_0$ and $H_1$ respectively the diagonal and off-diagonal parts of the original Hamiltonian. By definition $\Delta H_i=H_i-\inp{z}{H_i}{z}$ and so $f$ is defined on the domain pictured in figure \ref{domainfig}. The initial condition becomes  
\begin{equation}
f(\ave{H_0},\ave{H_1},\ell=0)=\ave{H_0}+\beta \ave{H_1}.
\end{equation}
Due to the coherent nature of the state (\ref{cohstate}) one might expect that terms corresponding to high order fluctuations in (\ref{lipkinhexp}) will contribute less significantly to $\inp{z}{H(\ell)}{z}$ than the scalar term $f$. This statement can be made precise as follows. Since the flowing Hamiltonian is extensive $f$ should be proportional to $j$, as is the case with $\ave{H_0}$ and $\ave{H_1}$. To keep track of the orders of $j$ we introduce the scaleless variables $x\equiv\ave{H_0}/j$, $y\equiv\ave{H_1}/j$ and $\tilde{f}\equiv f/j$. When taking the inner-product with respect to $\ket{z}$ on both sides of (\ref{lipkinhexp}) the linear terms fall away and we obtain
\begin{eqnarray}
	\fl \ave{H(\ell)}=f+\frac{1}{2}f^{11}\ave{\left\{\Delta H_0,\Delta H_1\right\}}+\frac{1}{2}f^{20}\ave{\Delta H_0^2}+\frac{1}{2}f^{02}\ave{\Delta H_1^2}+\cdots.
\end{eqnarray}
Each term is, up to a constant factor, of the form 
\begin{equation}
	f^{ik}\ave{{\rm Prod}(i+k)}=j^{1-(i+k)}\frac{\partial^{\,i+k} \tilde{f}}{\partial x^i\partial y^k}\ave{{\rm Prod}(i+k)}
\end{equation}
where ${\rm Prod}(i+k)$ denotes some arbitrary product of $i+k$ fluctuations. Using the results of \ref{appc} we see that such a term is at most of order $O(j^{1-(i+k)}j^{\left\lfloor (i+k)/2 \right\rfloor})$ in $j$. The leading order term corresponds to $i+k=0$, i.e. the scalar term $f$. We conclude that $f$ is the leading order contribution to $\inp{z}{H(\ell)}{z}$, expressed not as a function of $z$ and $z^*$, but rather of the averages $\ave{H_0}$ and $\ave{H_1}$. 

\subsection{The flow equation in the $j\rightarrow\infty$ limit.}
\label{flowinf}
We consider the flow of two operators, the Hamiltonian and another arbitrary operator $P$ which flows as $P(\ell)=U(\ell)^\dag P U(\ell)$. Since $H(\ell)$ determines $U(\ell)$ one would expect a one-way coupling between the equations. It is assumed that $P$ is a hermitian operator constructed in terms of $H_0$ and $H_1$. Furthermore $\inp{z}{P}{z}$ must be a rational function of $j$, which ensures that when taking derivatives of $g(x,y,\ell)$ no additional factors of $j$ are generated. First we summarize the equations concerned
\begin{eqnarray}
		\fl H(\ell)=f+f^{10}\Delta H_0+f^{01}\Delta H_1+\frac{1}{2}f^{11}\left\{\Delta H_0,\Delta H_1\right\}+\frac{1}{2}f^{20}\Delta H_0^2+\frac{1}{2}f^{02}\Delta H_1^2+\cdots,\label{expand2}\\
		\fl P(\ell)=g+g^{10}\Delta H_0+g^{01}\Delta H_1+\frac{1}{2}g^{11}\left\{\Delta H_0,\Delta H_1\right\}+\frac{1}{2}g^{20}\Delta H_0^2+\frac{1}{2}g^{02}\Delta H_1^2+\cdots,\\
		\fl \frac{d H(\ell)}{d\ell}=\comm{\comm{H_0}{H(\ell)}}{H(\ell)},\ \ \ H(0)=H_0+\beta H_1, \label{floweqh2} \\
		\fl \frac{d P(\ell)}{d\ell}=\comm{\comm{H_0}{H(\ell)}}{P(\ell)},\ \ \ P(0)=P,
\end{eqnarray}
where $f(\ave{H_0},\ave{H_1},0)=\ave{H_0}+\beta\ave{H_1}$ and $g(\ave{H_0},\ave{H_1},0)$ is just $\inp{z}{P}{z}$ up to leading order in $j$. Next we substitute the expansion of $H(\ell)$ into the flow equation (\ref{floweqh2}) and take the expectation value on both sides with respect to the state $\ket{z}$. Arguing as before we identify the leading order term on the left as being ${\rm \partial}f/{\rm \partial}{\ell}$. A general term on the right is of the form
\begin{equation}
	f^{in}f^{kl}\ave{\comm{\comm{\Delta H_0}{{\rm Prod}(i+n)}}{{\rm Prod}(k+l)}}.
\end{equation}
By transforming to scaleless variables and using the result of \ref{appc} it is seen to be at most of order $O(j^{2-t+\left\lfloor(1+t)/2\right\rfloor})$, where $t=i+n+k+l$. Since the $t=0$ and $t=1$ terms are zero (they involve commutators of scalars) the leading order contributions come from the $t=2,3$ terms. These are exactly the terms found by considering the expansion of $H(\ell)$ up to second order in the fluctuations. The expectation values of the double commutators may be calculated and expressed as functions of $x$ and $y$ using the method outlined in \ref{appb}.
It is found that, due to cancellations, only four of the potential fifteen terms make leading order contributions. The corresponding expectation values are, to leading order
\begin{eqnarray}
	\ave{\comm{\comm{\Delta H_0}{\Delta H_1}}{\Delta H_0}}&=& g_{21}(x,y)j\\
	\ave{\comm{\comm{\Delta H_0}{\Delta H_1}}{\Delta H_1}}&=&g_{22}(x,y)j\\
	\ave{\comm{\comm{\Delta H_0}{\acomm{\Delta H_0}{\Delta H_1}}}{\Delta H_1}}&=&g_{32}(x,y)j^2\\
	\ave{\comm{\comm{\Delta H_0}{\Delta H_1^2}}{\Delta H_0}}&=&-g_{32}(x,y)j^2
\end{eqnarray}
where $g_{21}=-4y$, $g_{22}=2x(1-x^2)$, $g_{32}=-2+4x^2-2x^4+8y^2$. When keeping only the leading order terms on both sides the flow equation becomes
\begin{equation}
\pafg{\tilde{f}}{\ell}=g_{21}\tilde{f}^{01}\tilde{f}^{10}+g_{22}\tilde{f}^{01}\tilde{f}^{01}+\frac{1}{2}g_{32}\tilde{f}^{11}\tilde{f}^{01}-\frac{1}{2}g_{32}\tilde{f}^{02}\tilde{f}^{10},
\end{equation}
where superscripts denote derivatives to the rescaled averages $x$ and $y$. For further discussion it is convenient to transform to the variables $(p=r^2,q=\cos(2\theta))\in\left[0,\infty\right)\times\left[-1,1\right]$, which are related to $(x,y)$ by
\begin{equation}
	x=\frac{p-1}{p+1}\ \ \ {\rm and}\ \ \ y=\frac{2pq}{(1+p)^2}+O(\frac{1}{j}).	
\end{equation}
Note that the domain of these variables is a strip, which significantly simplifies the numerical solution. We note that the arguments above can be applied, completely unchanged, to the flow equation of $P(\ell)$ as well. We obtain, now for both $H(\ell)$ and $P(\ell)$, the coupled set
\begin{eqnarray}
\pafg{\tilde{f}}{\ell}=-2(1+p)^2\left(q\frac{\partial\tilde{f}}{\partial q}\frac{\partial\tilde{f}}{\partial p}+(1-q^2)\frac{\partial\tilde{f}}{\partial q}\frac{\partial^2\tilde{f}}{\partial q\partial p}\right) \label{couplesetforH}\\
\pafg{g}{\ell}=-2(1+p)^2\left(q\frac{\partial\tilde{f}}{\partial q}\frac{\partial g}{\partial p}+(1-q^2)\frac{\partial g}{\partial q}\frac{\partial^2\tilde{f}}{\partial q\partial p}\right).
\label{couplesetforHandN}
\end{eqnarray}
Note that in contrast to the equation for $\tilde f$, the equation for $g$ is a linear equation that can be solved once $\tilde f$ has been obtained from (\ref{couplesetforH}).
 
In section \ref{lipkinflow} it was mentioned that $H(\ell)$ retains its band diagonal structure during flow, which meant that $H_1$ appeared only linearly in the representation of $H(\ell)$. This implies that $f(\ave{H_0},\ave{H_1},\ell)$ should be linear in $\ave{H_1}$, or, in the new variables, linear in $q$. When the form $\tilde{f}(p,q,\ell)=n_0(p,\ell)+q n_1(p,\ell)$ is substituted into (\ref{couplesetforH}) this is indeed seen to be the case, and we obtain a remarkably simple set of coupled PDE's for $n_0$ and $n_1$:
\begin{eqnarray}
	\frac{\partial n_0}{\partial \ell}=-2(1+p)^2n_1\frac{\partial n_1}{\partial p}\nonumber \\
	\frac{\partial n_1}{\partial \ell}=-2(1+p)^2n_1\frac{\partial n_0}{\partial p}.
	\label{simplesetcoupled}
\end{eqnarray}
The initial conditions are 
\begin{equation}
	n_0(p,0)=x=\frac{p-1}{p+1}\ \ \ {\rm and}\ \ \ n_1(p,0)=\frac{\beta y}{q}=\frac{2p\beta}{(p+1)^2}.
\end{equation}
The matrix elements of the Lipkin Hamiltonian possess the symmetry $\inp{m}{H}{m}=-\inp{-m}{H}{-m}$, which implies that the spectrum is anti-symmetric around $E_j=0$. This symmetry is respected by the flow equation and manifests itself through an invariance of equations (\ref{simplesetcoupled}) under the substitutions $n_0\rightarrow-n_0$, $n_1\rightarrow n_1$ and $p\rightarrow p^{-1}$. For the solutions this implies that $n_0(p,\ell)=-n_0(p^{-1},\ell)$ and $n_1(p,\ell)=n_1(p^{-1},\ell)$, and we may thus restrict ourselves to the interval $[0,1]$ in the $p$ dimension. This will be seen to correspond to the negative half of the spectrum, from which the entire spectrum can easily be obtained. Apart from the initial conditions we impose the boundary conditions $n_0(1,\ell)=0$, as required by symmetry considerations, and $n_1(0,\ell)=0$, since at $r=0$ the $\theta$ dependency should vanish. The latter is also consistent with and, in fact, follows from the flow equation and initial value at $p=0$, since $\partial n_1/\partial \ell \propto n_1$. A naive application of the same argument to $n_0$ seems to suggest that no flow occurs for $n_0$ at $p=0$ and that $n_0(0,\ell)=-1$ for all $\ell$. This conclusion is, however, incorrect since a more careful analysis shows that in the second phase $\partial n_1/\partial p$ develops a square root singularity at $p=0$, which allows $n_0(0,\ell)$ to flow away from $-1$. Numerically this can be handled easily by solving for $m(p,\ell)\equiv n_1^2(p,\ell)$, instead of $n_1$. 
\subsection{Finding the spectrum}
\label{spectrum}
From $f=j\tilde{f}$ it is possible to obtain the entire spectrum of $H$. For this discussion it is convenient to use the original $(\ave{H_0},\ave{H_1})$ variables. Recall that in the $l\rightarrow\infty$ limit $H(\ell)$ flows toward a diagonal form, furthermore the eigenvalues appear on the diagonal in the same order as in $H_0$, i.e. increasing from top to bottom. This implies that the $n^{\rm th}$ eigenvalue of $H$ is given by
\begin{equation}
	E_n=\inp{-j+n}{H(\ell=\infty)}{-j+n},
\end{equation}
where $E_0$ corresponds to the ground state energy. As $H(\ell)$ flows towards a diagonal form the terms of expansion (\ref{expand2}) containing $H_1$ will disappear and eventually $H(\infty)$ and $f$ will become functions of $H_0$ and $\ave{H_0}$ only. The eigenvalues of $H$ are given by
\begin{equation}
	E_n=f(\ave{H_0}=-j+n,0,\ell=\infty),
\end{equation}
where $\ave{H_1}$ has been set to zero. This can be understood in two ways. Looking at equations (\ref{gammabardef}) and (\ref{genparamh}) we see that the functional dependence of $H(\ell)$ on $H_0$ is the same as that of $\bar{\Gamma}$ (now called $f$) on $\ave{H_0}$. Since taking the expectation value of $H(\infty)$ with respect to $\ket{-j+n}$ is equivalent to substituting $-j+n$ for  $H_0$, the result follows. Alternatively, consider equation (\ref{fexpandflux}) and note that setting $\ave{H_0}=-j+n$ makes the ($n+1)^{\rm st}$ diagonal element of $\Delta H_0$ zero. Since $H_1$ does not appear we see that when taking the inner product with $\ket{-j+n}$ only the scaler term, $f$, will survive. The results of this procedure appear in section \ref{numspectrum}.
\subsection{Calculating expectation values}
\label{expval}
Next we return to the arbitrary operator $P$ introduced earlier, and show how its expectation values with respect to the eigenstates of $H$ may be calculated. As examples we consider the first and second moments of $H_0=J_z$ in the ground state. The former is simply related to the order parameter $\Omega$ discussed in section \ref{lipkinmodel}. Suppose $P$ is a hermitian operator having the properties mentioned in section \ref{flowinf}. The aim is to calculate $\inp{E_n}{P}{E_n}$ where $\ket{E_n}$ is the (unknown) eigenstate of $H=H_0+\beta H_1$ corresponding to the energy $E_n$. We can formulate this calculation in terms of a flow equation by noting that
\begin{equation}
	\inp{E_n}{P}{E_n}=\inp{E_n}{U(\ell)U^\dag(\ell) PU(\ell)U^\dag(\ell)}{E_n}=\inp{E_n,\ell}{P(\ell)}{E_n,\ell}
\end{equation}
where $\ket{E_n,\ell}=U^\dag(\ell)\ket{E_n}$, $P(\ell)=U^\dag(\ell)PU(\ell)$ and $U(\ell)$ is the unitary operator associated with the flow of $H(\ell)$. This equation holds for all $\ell$, and particularly in the $\ell\rightarrow\infty$ limit. Since $\ket{E_n,\infty}=\ket{-j+n}$ it follows that 
\begin{equation}
	\inp{E_n}{P}{E_n}=\inp{-j+n}{P(\infty)}{-j+n},
\end{equation}
and we conclude that the expectation value of $P$ in the $n^{\rm th}$ eigenstate of $H$ is simply the $n^{\rm th}$ diagonal element of $P(\infty)$ in the $H_0$ basis. Furthermore $\inp{z}{P(\ell)}{z}$ has the character of a generating function in the sense that
\begin{equation}
	\left.\frac{\partial^{n+m}\inp{z}{P(\ell)}{z}}{{\partial z^*}^n\partial z^m}\right|_{z=z^*=0}\propto\inp{-j+n}{P(\ell)}{-j+m},
\end{equation}
thus knowing $g$, which is just $\inp{z}{P(\ell)}{z}$ up to leading order in $j$, is sufficient to obtain all the matrix elements of $P(\ell)$ with high accuracy. However, unless an analytic solution for $g$ is known, we are limited to numerical calculations for the low lying states. In particular, the ground state expectation value is found by setting $z=0$. 

Consider the case where $P=H_0/j=J_z/j$ and $g(x,y,\ell=0)=x$. Here $P(\ell)$ will flow from a diagonal form to one containing high powers of $H_1$. Having no knowledge about the form of $g$ we must solve the flow equation directly as a linear PDE in 3 variables. For numerical implementation we again change to the variables $p$, $q$ and solve on the strip $p\in [0,1]$, $q\in [-1,1]$.  As before we impose the boundary condition $(\frac{\partial g}{\partial q})_{p=0}=0$.  It turns out that the solution for $p<<1$ is virtually independent of the boundary condition imposed at $p=1$ (this is in fact also true for $f$ or $n_0$ and $n_1$) and in this case we simply allow the solution to evolve freely at $p=1$. The results of this calculation is shown in section \ref{numorder}.
\subsection{Probing the structure of eigenstates}
\label{eigenstates}
The ability to calculate expectation values for the class of operators discussed above allows us to probe the eigenstates in a variety of ways. We will consider diagonal operators of the form $O(h)=\sum_{n=-j}^{j}h(n/j)\ket{n}\bra{n}$ where $h$ is a smooth, $j$-independent function defined on $[-1,1]$. Since $H_0$ is non-degenerate its powers span the space of diagonal operators, and so $O(h)$ can always be expressed as a function of $H_0$. In order to apply the flow equation to $O(h)$ we must find the initial condition for equation (\ref{couplesetforHandN}), which is given by
\begin{equation}
	g(x,y,\ell=0)=\lim_{j\rightarrow\infty}\inp{z}{O(h)}{z}.
\end{equation}
The expansion coefficients for the coherent state in the $J_z$ basis are given by \cite{Perelomov}
\begin{equation}
	\braket{n}{z}=(1+r^2)^{-j}{ 2j \choose n+j}^{1/2}z^{n+j}\ \ \ \ \ n=-j,\ldots,j
\end{equation}
which, together with a change of variables from $r$ to $x=(r^2-1)/(r^2+1)$, lead to
\begin{equation}
	\inp{z}{O(h)}{z}=\sum_{n=0}^{2j}h\left(-1+\frac{n}{j}\right){ 2j \choose n}\left(\frac{1-x}{2}\right)^{2j-n}\left(\frac{x+1}{2}\right)^{n}.
\end{equation}
This is recognized as the Bernstein polynomial \cite{Davis} for $h(x)$ of degree $2j$ on the interval $[-1,1]$. It is well known that in the $j\rightarrow\infty$ limit this polynomial approximation converges uniformly to $h(x)$, from which we conclude that the boundary condition of the flow equation is simply $g(x,y,\ell=0)=h(x)$. \\

Suppose $\ket{E_m}=\sum_{n=-j}^{j}\alpha_n\ket{n}$ is the eigenstate under consideration, in which case
\begin{equation}
	\inp{E_m}{O(h)}{E_m}=\sum_{n=-j}^{j}h\left(\frac{n}{j}\right)\left|\alpha_n\right|^2=\sum_{n=-j}^{j}h\left(\frac{n}{j}\right)\left|\braket{n}{E_m}\right|^2.
	\label{eigenstateinp}
\end{equation}
This leads to the interpretation of the expectation value as an average of the expansion coefficients $|\braket{n}{E_m}|^2$, weighted by the function $h(x)$. It may appear that choosing $O(h)$ to be the projection operator onto some basis state $\ket{n}$ provides the most direct means of calculating the contributions of individual states. However the projection operator does not fall within the class of operators for which equation (\ref{couplesetforHandN}) was derived. 
This follows from the fact that when taking derivatives of $\braket{z}{n}\braket{n}{z}$ new factors of $j$ are generated, which is equivalent to the observation that no continuous function $h(x)$ exists such that $O(h(x))=\ket{n}\bra{n}$ for all $j$. Instead we choose $h(x,\bar{x})=\exp(-\gamma(\bar{x}-x)^2)$ where $\bar{x}\in[-1,1]$ and $\gamma>>1$. This weight function focuses on the contribution of those basis states $\ket{n}$ for which $n/j$ lies in a narrow region centered around $\bar{x}$. Considered as functions of $\bar{x}$, it is expected that $\ave{O(h(x,\bar{x}))}$ would approximate $\lim_{j\rightarrow\infty}\left|\alpha_{j\bar{x}}\right|^2$ up to a constant factor, provided that the latter varies slowly on the scale of $1/\gamma$ in the region of $\bar{x}$. Clearly $\gamma$ controls the accuracy of this method, and also determines the scale on which the structure of the eigenstates can be resolved. 

\section{Numerical Results}
\label{numres}
\begin{figure}[ht]
\begin{tabular}{cc}
\epsfig{file=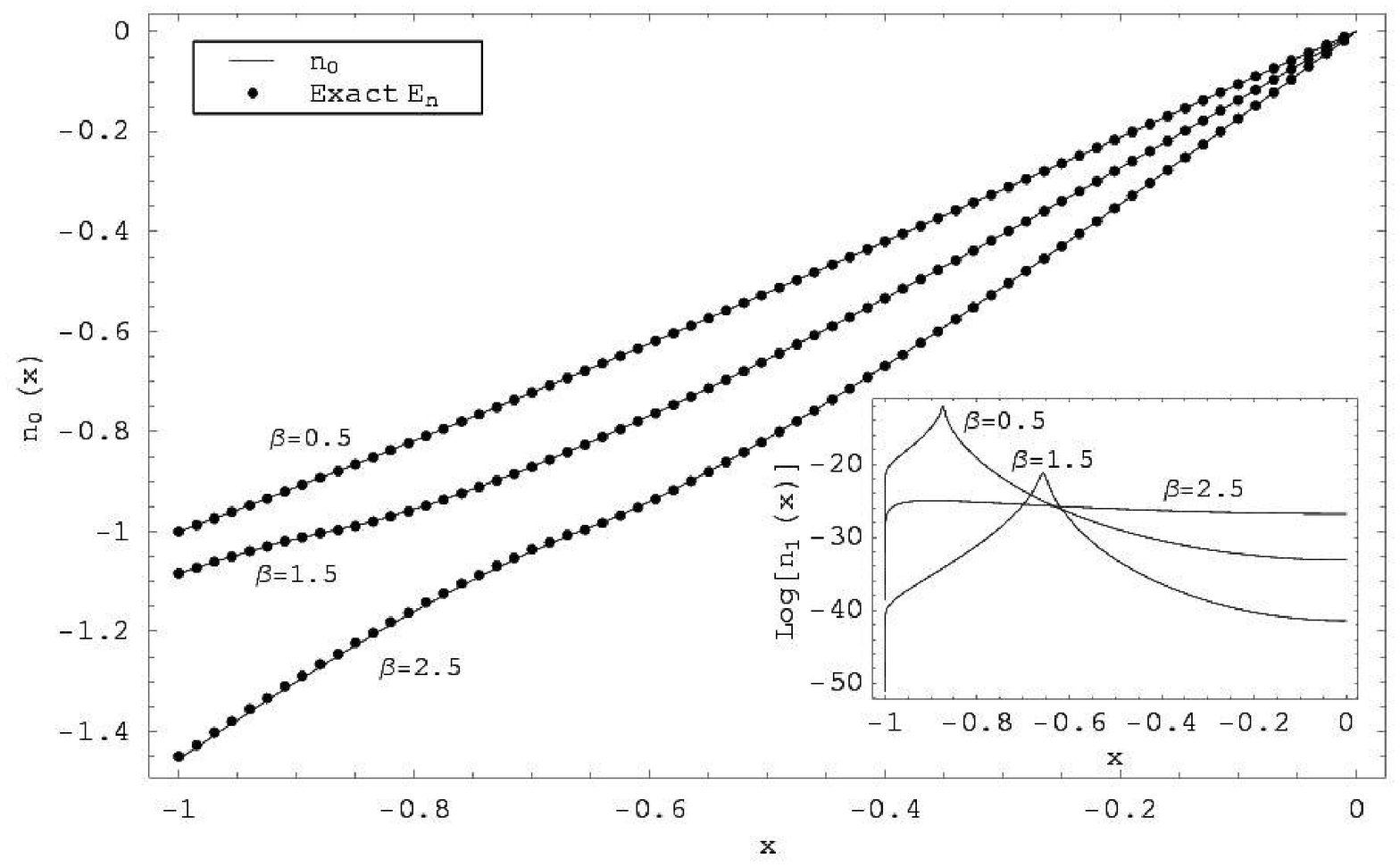,height=5cm,clip=,angle=0} &
 \epsfig{file=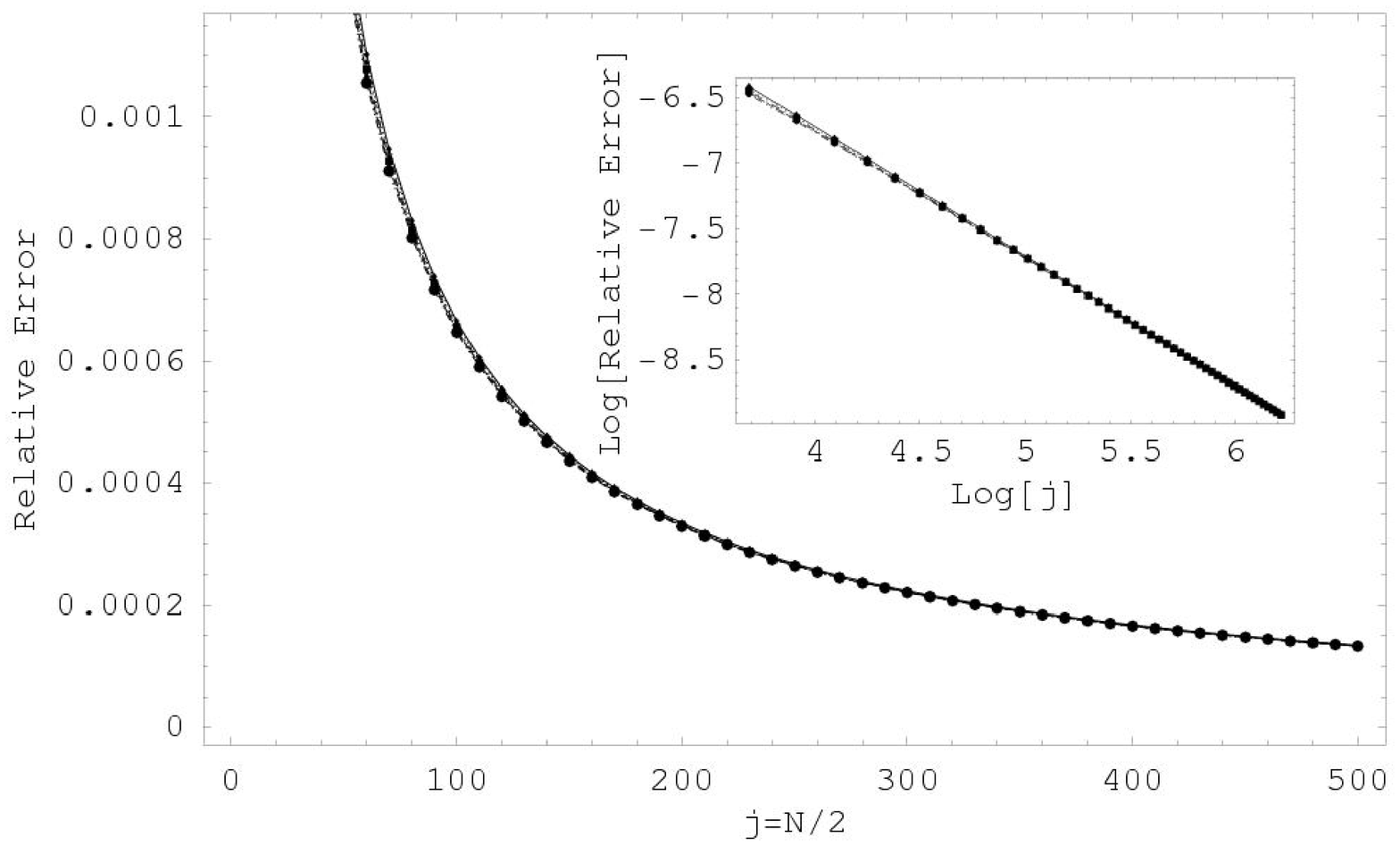,height=5cm,clip=,angle=0} \\
	(a) & (b)\\
	\vspace{0.05cm} &\\
\epsfig{file=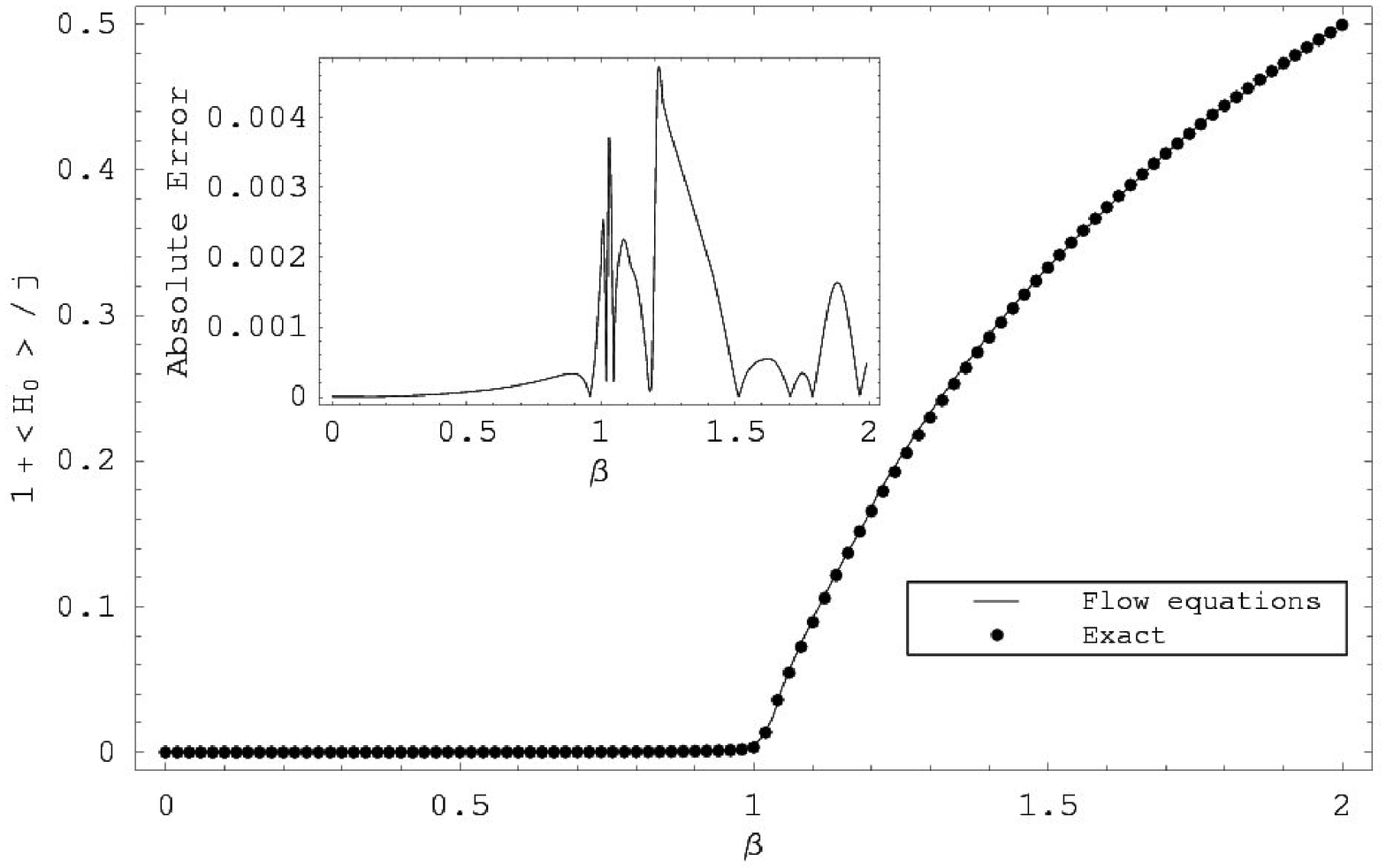,height=5cm,clip=,angle=0} & \epsfig{file=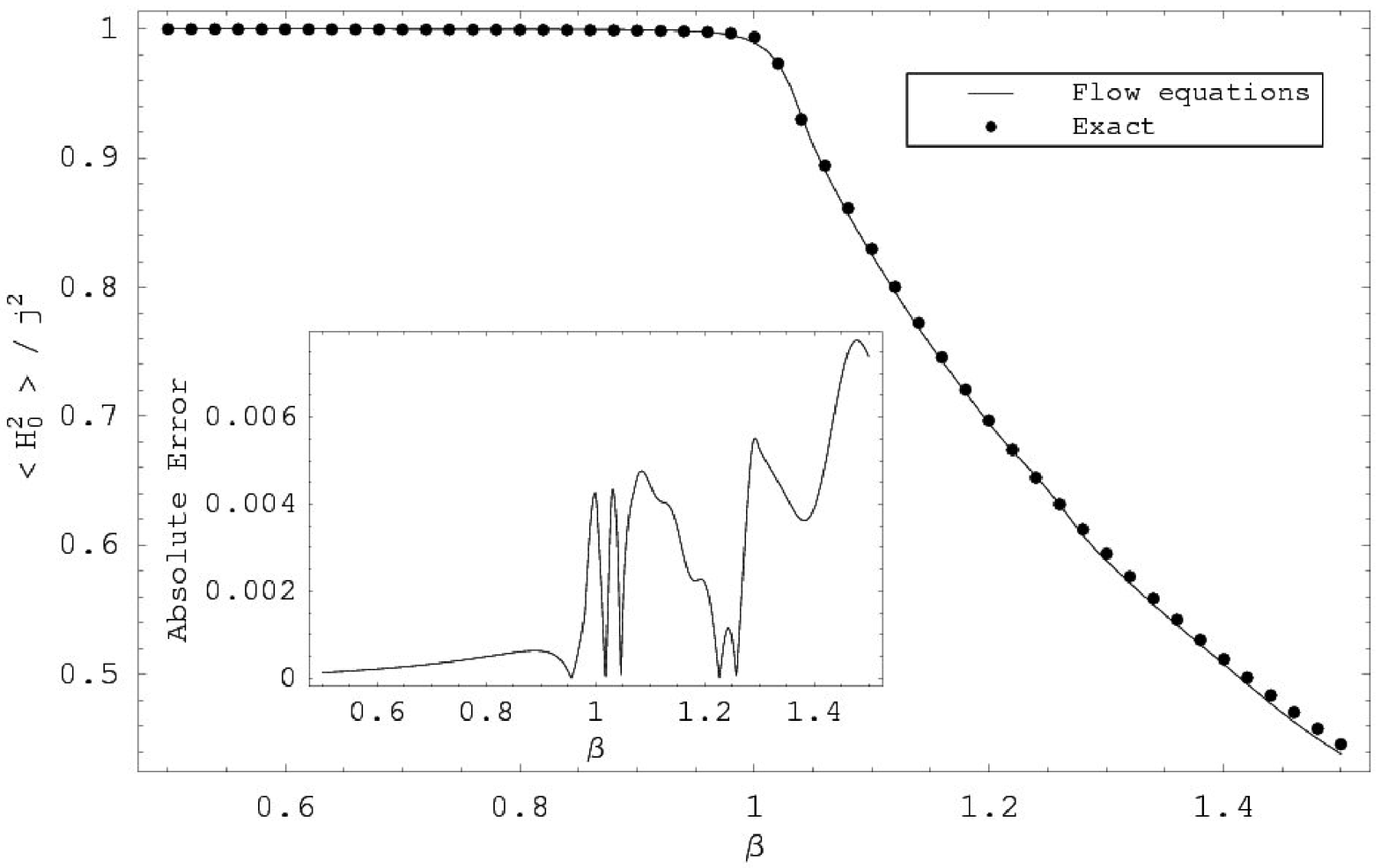,height=5cm,clip=,angle=0}  \\
	(c) & (d)
\end{tabular}
	\caption[gr1]{Results obtained from the numerical solution of the flow equations. (a) The full spectrum.  For clarity the exact result is only displayed for every fifteenth eigenvalue. (b) Relative error in the first five states as a function of $j$ for $\beta=0.5$.  (c) The order parameter $\Omega=1+\ave{H_0}/j$ (in the ground state) as a function of $\beta$.  (d) $\ave{H_0}^2/j^2$ in the ground state as a function of $\beta$. }
	\label{gr1}
\end{figure}

\begin{figure}[ht]
\begin{tabular}{cc}
	\epsfig{file=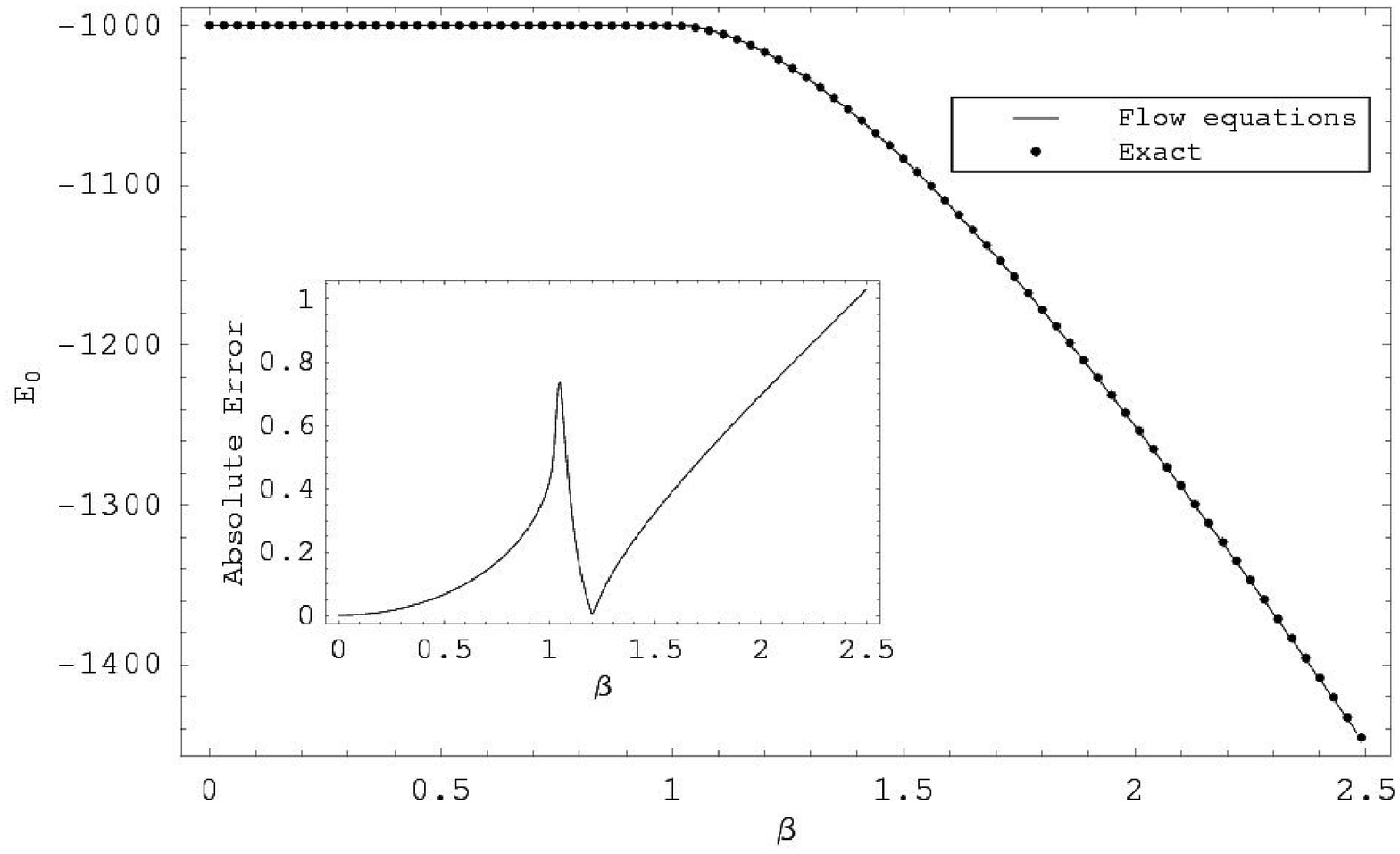,height=5cm,clip=,angle=0} &
	\epsfig{file=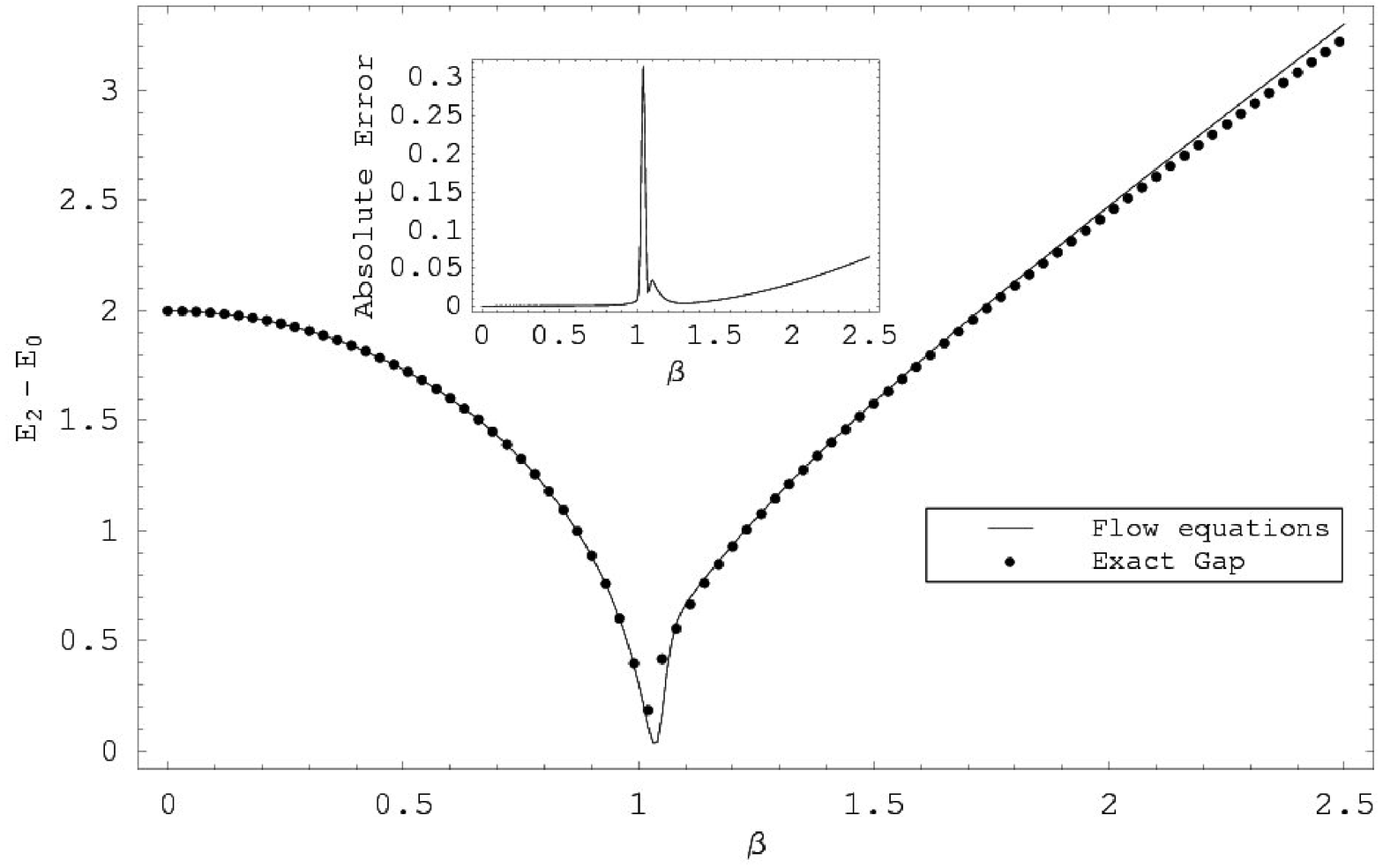,height=5cm,clip=,angle=0} \\
	(a) & (b)\\
	\vspace{0.05cm} &\\
	\epsfig{file=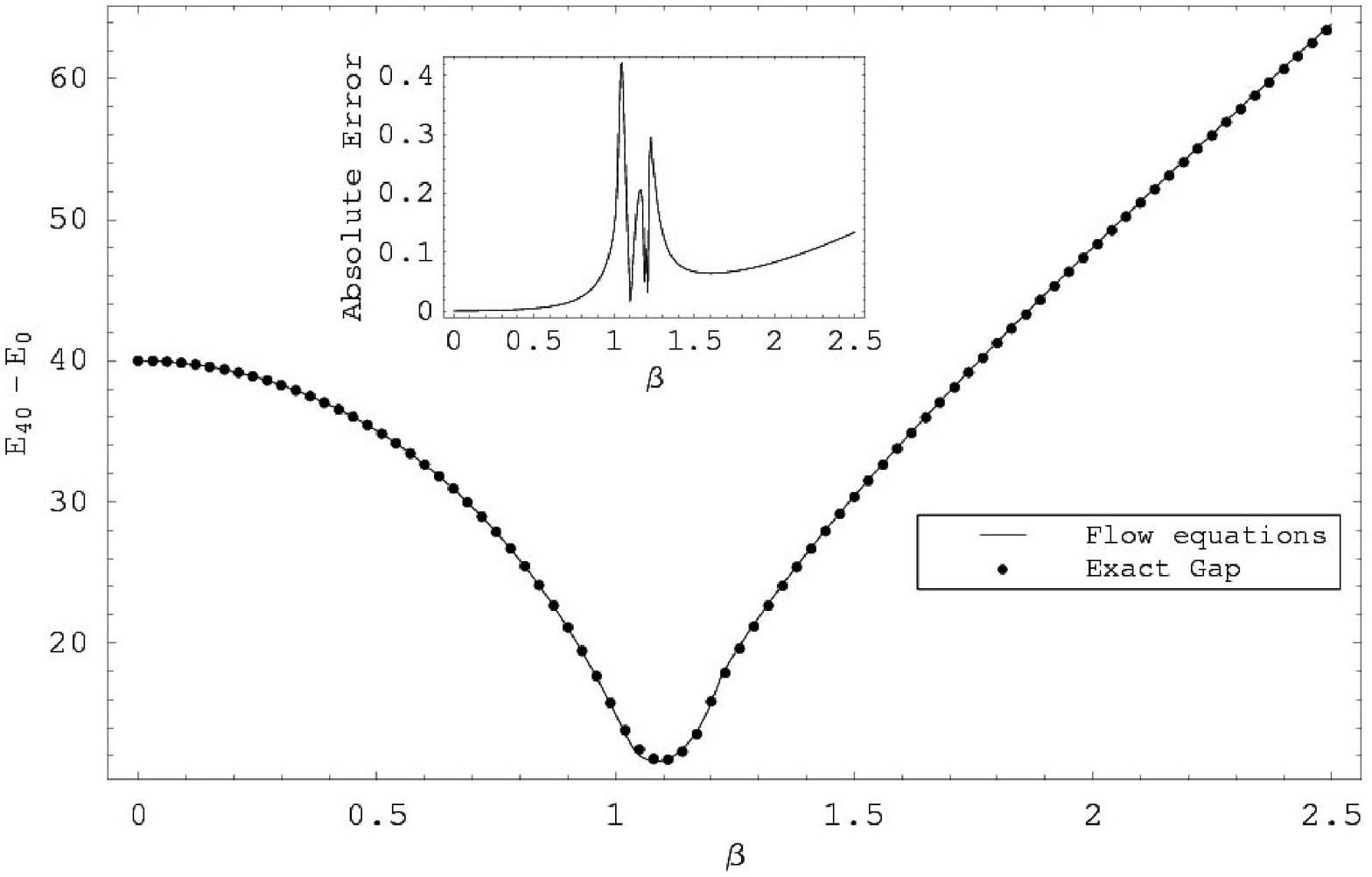,height=5cm,clip=,angle=0} &
	\epsfig{file=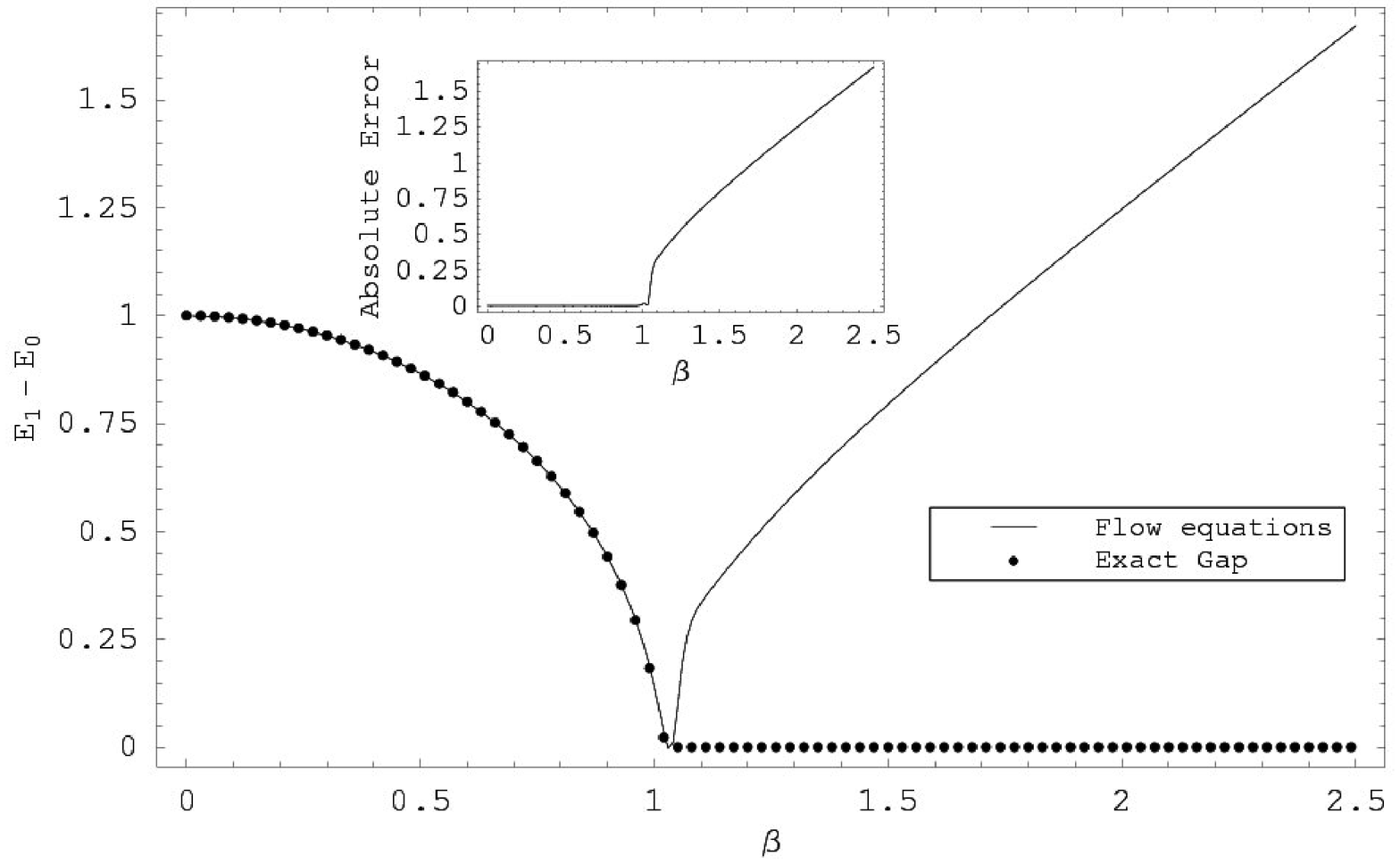,height=5cm,clip=,angle=0}  \\
	(c) & (d)
\end{tabular}
	\caption[gr2]{Energy levels and gaps as functions of the coupling constant.(a) The ground state energy.  (b) The gap between the second excited state and ground state. (c) The gap between the fortieth excited state and ground state.  (d) Gap between the first excited state and ground state.}
	\label{gr2}
\end{figure}

\subsection{The flow of $H(l)$}
\label{numspectrum}
In section \ref{spectrum} it was shown the ($n+1)^{\rm st}$ eigenvalue of $H$ is given by $E_n=f\left(-j+n\right)$, where we suppress the last two arguments and take $\ave{H_1}=0$ and $\ell=\infty$ in what follows. Note that $f$ may be interpreted as a mapping of the non-interacting spectrum onto the interacting spectrum of $H$ at finite $\beta$. In terms of the rescaled variables this becomes 
\begin{equation}
	E_n=j \tilde{f}\left(x=-1+n/j\right)=j n_0\left(p=\frac{1+x}{1-x}\right)\equiv j n_0\left(x\right),
\end{equation}
where we use $n_1(p,\ell=\infty)=0$. As $j\rightarrow\infty$ the difference between values of $x$ corresponding to successive states become zero, and we may consider $x\in[-1,1]$ as a continuous label for the eigenstates through the assosiation $x\leftrightarrow\ket{E_{j(1+x)}}$.

Another property of interest is the gap between successive levels, which, for large $j$, is given by
\begin{eqnarray}
	E_{n+1}-E_{n}&=&j\left[n_0\left(x=-1+\frac{n+1}{j}\right)-n_0\left(x=-1+\frac{n}{j}\right)\right]\nonumber\\
	&\approx&\left.\frac{\partial n_0}{\partial x}\right|_{x=-1+n/j}.
	\label{gapasdif}
\end{eqnarray}
Figure \ref{gr1} (a) shows the numerical solution for $n_0(x)$ at $\ell=6$ and for three different values of $\beta$. The inset is the corresponding graph for $\log(n_1)$. For these values of the flow parameter $n_1$, which represents the off-diagonal part of $H(\ell)$, is already of order $10^{-20}$ and may be disregarded completely. As a first test we calculate the spectrum for $j=1000$ from direct diagonalization and plot the pairs $(x=-1+n/j,E^{exact}_n/j)$ for $n=0,\ldots,j$ as dots. There is an excellent correspondence for all states and values of the coupling constant as reflected in figure \ref{gr1} (b) which shows the relative error in the first five energy eigenvalues as a function of $j$ for $\beta=0.5$. The lines fall almost exactly on one another, so no legend is given. The log-log inset shows a set of straight lines with gradients equal to one, clearly illustrating the $1/j$ behaviour of the errors. The same results hold in the second phase, although here the eigenvalues corresponding to the even states $n=0,2,4,\ldots$ tend to be much more accurate than those for odd states.  We return to this point shortly.  

Next we investigate how the eigenvalues depend on the coupling $\beta$. Figure \ref{gr2} (a) shows the ground state energy as a function of $\beta$ for $j=1000$ together with the exact result. The relative error is about $0.08\%$. As one would expect from the global spectrum in figure \ref{gr1} the gaps are produced to a good accuracy.  As examples we display the gaps $E_2-E_0$ and $E_{40}-E_0$ in figures \ref{gr2} (b) and (c). The absolute error is shown as an inset. 

One quantity that is not correctly reproduced is the gap between the ground state and first excited state. Figure \ref{gr2} (d) displays this quantity, which turns out to be correct only in the first phase.  Exact diagonalization shows that this gap vanishes in the second phase as $1/j$, while the flow equation produces a gap growing linearly with $\beta$.  The origin of this problem is, however, well understood and relates to the observation made above regarding the accuracy of the even states $n=0,2,4,\ldots$ versus the odd states in the second phase.  The reason why the flow equations, at least as implemented here, fail to reproduce this quantity correctly relates to the lack of irreducibilty of $\{H_0,H_1\}$ on the full space (see section \ref{flowrep}).  In principle one should implement the flow equations on each invariant subspace $\left\{\ket{n}\,|\,n\ {\rm even}\right\}$ and $\left\{\ket{n}\,|\,n\ {\rm odd}\right\}$ seperately, in contrast to what has been done here.  This will produce two functions, mapping the non-interacting spectrum in each subspace onto the corresponding interacting spectrum.  In this way the correct features of both sets of excitations can be reproduced.  This demonstrates that care should be taken to ensure that the underlying symmetries and associated selection rules of the Hamiltonian are properly taken into account in the flow equation.\\

\begin{figure}[ht]
\begin{tabular}{cc}
	\epsfig{file=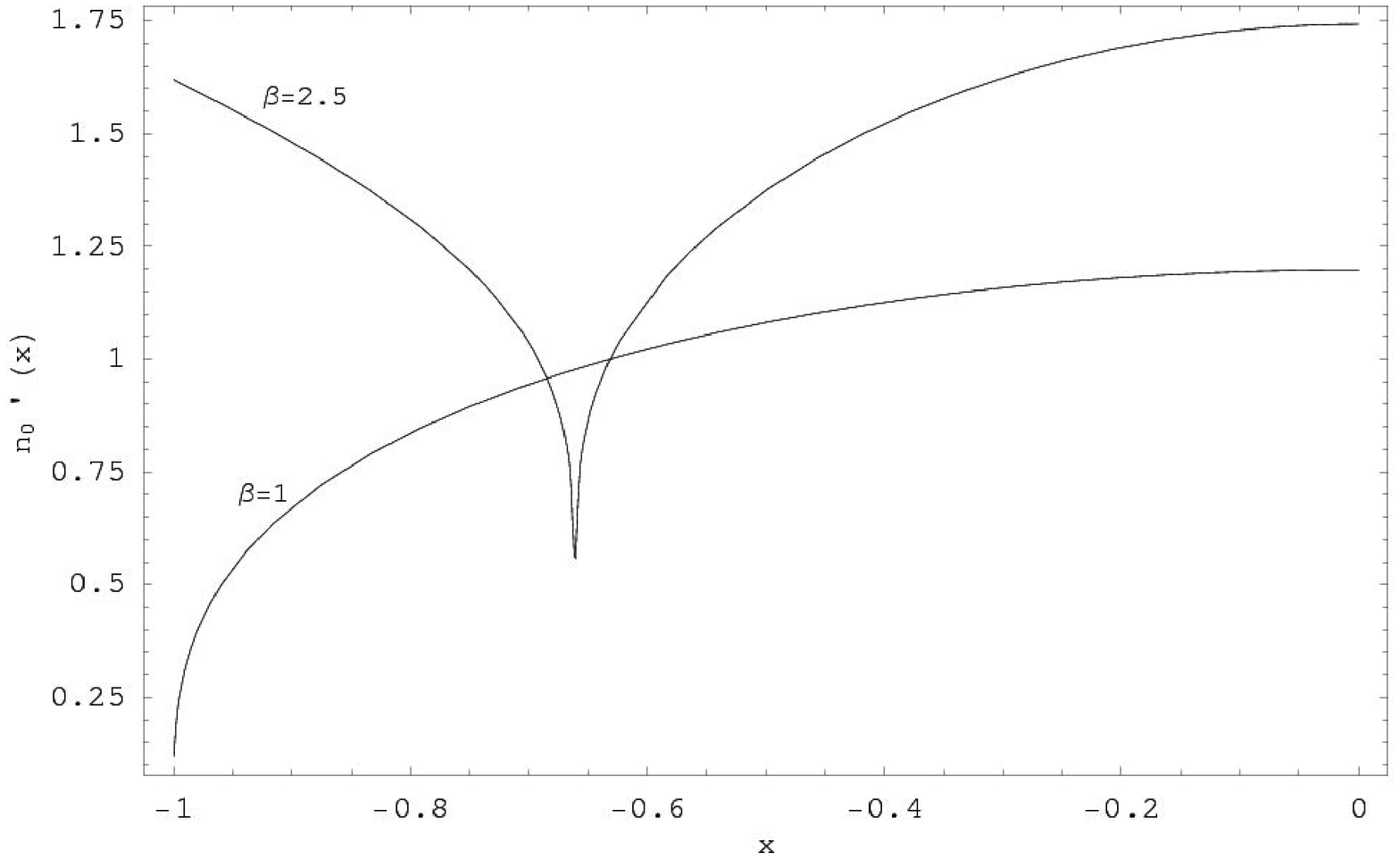,height=5cm,clip=,angle=0} &
	\epsfig{file=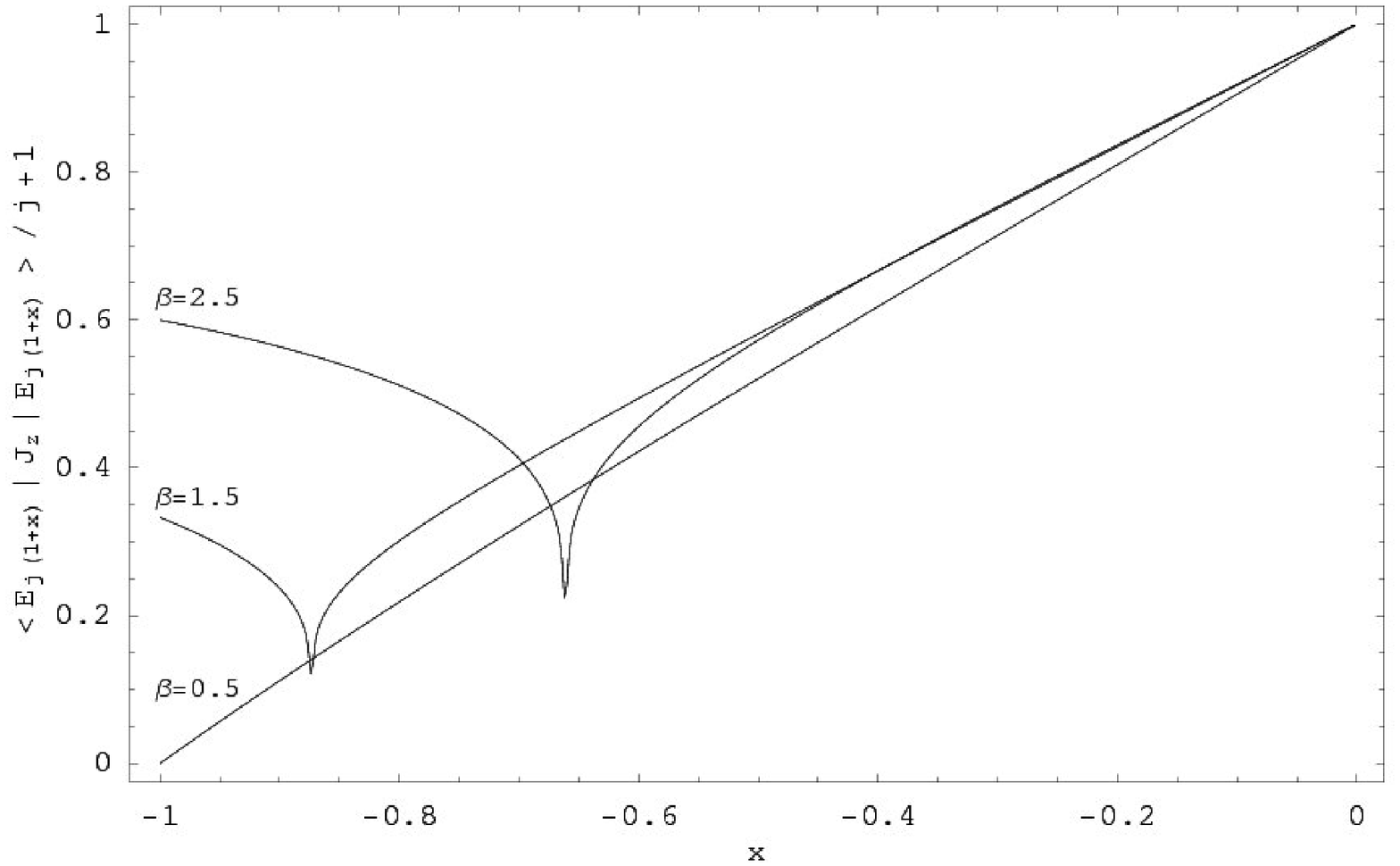,height=5cm,clip=,angle=0} \\
	(a) & (b)\\
	\vspace{0.05cm} &\\
	\epsfig{file=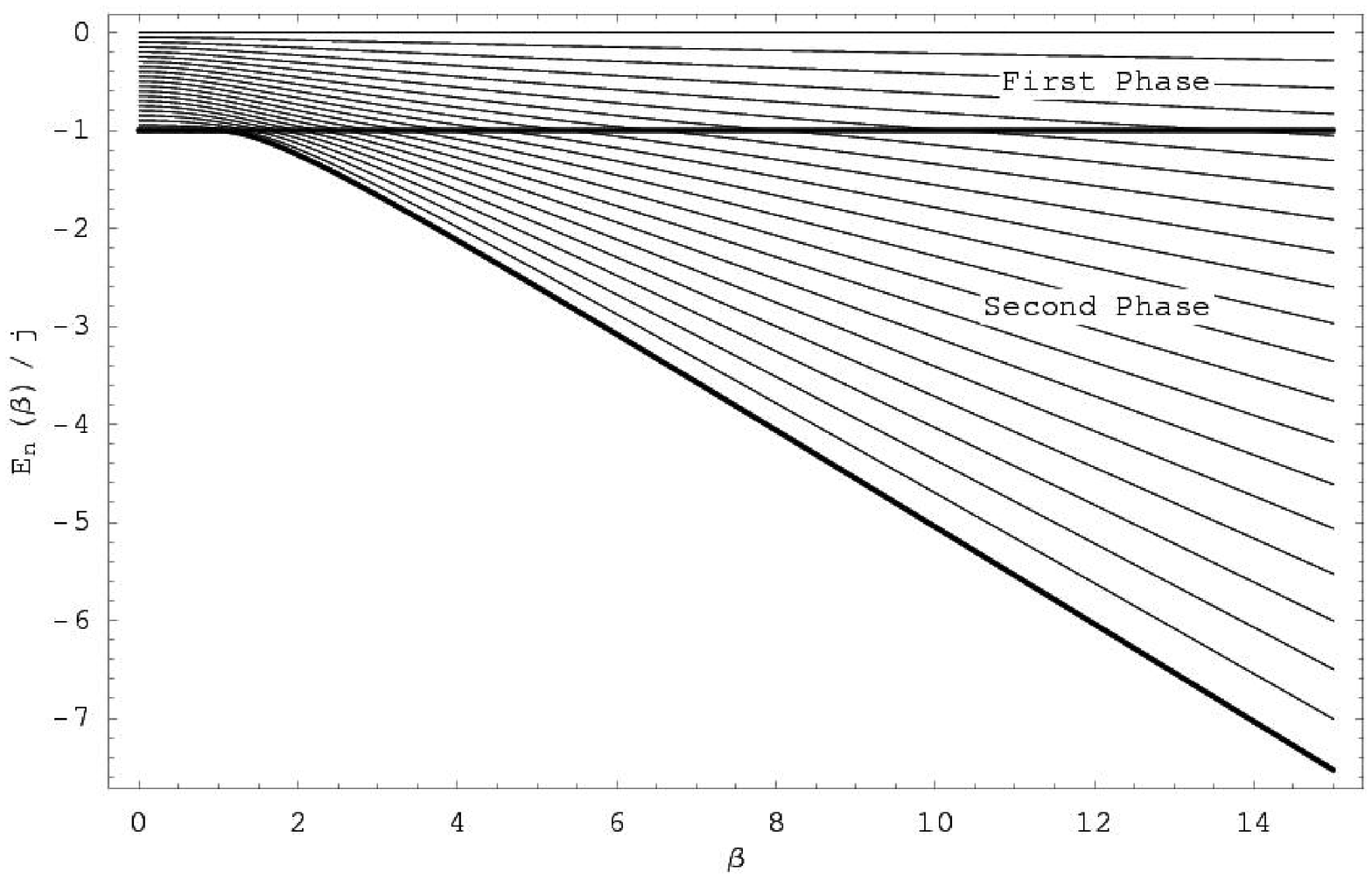,height=5cm,clip=,angle=0} &
	\\
	(c) 
\end{tabular}
	\caption[orderex]{(a) Derivative of $n_0(x)$ as a function of $x$ for $\beta=1, 2.5$. (b) $1+\inp{E_{j(1+x)}}{J_z}{E_{j(1+x)}}/j$ as a function of $x$ for $\beta=0.5, 1.5, 2.5$ and $j=1000$. (c) The negative half of the spectrum as a function of $\beta$ for $j=1000$. Only every twenty-fifth even eigenvalue is shown.}
	\label{orderex}
\end{figure}

The flow equation provides an interesting new perspective on the quantum phase transition which we now discuss. We mention that the graphs in figure \ref{orderex}, which illustrate properties of the Lipkin model not directly related to flow equations, were obtained from the direct diagonalization of $H$ for $j=1000$. Where applicable the flow equation gives the same results. As is well known \cite{Sachdev} a quantum phase transition is related to a large number of avoided level crossings.  This is clearly displayed in figures \ref{gr2} (b) and (c) where the approximate vanishing of the gap at the critical value $\beta=1$ can clearly be seen. From equation (\ref{gapasdif}) we note that a vanishing gap corresponds to a vanishing  derivative of $n_0(x)$.  From this point of view, avoided level crossings manifest themselves as a vanishing derivative of $n_0(x)$, the function that maps the non-interacting spectrum onto the interacting spectrum.  From figure \ref{gr1} (a) it is clear that such a vanishing derivative never occurs when $\beta<1$.  For values of $\beta\geq 1$, however, one always finds a point where the curve $n_0(x)$ forms a plateau and the derivative vanishes approximately as shown in figure \ref{orderex} (a). In fact at $\beta=1$ this happens right at $x=-1$, i.e., at the ground state energy so that at this value of $\beta$ the gap between the lowest excited states and the ground state vanishes, consistent with our earlier observation. At larger values of $\beta$ this point shifts up in the spectrum as is clear from figure \ref{gr1} (a). Note from figure \ref{gr1} (a) that at this point $n_0=-1$ for all values of $\beta$.  We remark that the approximate vanishing of the derivative is probably due to numerics.  If the derivative does vanish exactly at some point, which we believe to be the case, it will be very hard to detect it numerically as the flow equation is of course solved on a grid with finite spacing. Note that a vanishing derivative of $n_0(x)$ does not imply a vanishing gap at finite $j$ (number of particles) as $n_0(x)$ is then only evaluated at a finite number of discrete points, excluding to all probability the point where the derivative of $n_0(x)$ vanishes.  It must be kept in mind that $n_0(x)$ really reflects the situation in the thermodynamic limit and therefore it displays a point where the derivative vanishes.    

Returning to the flow equation $(\ref{simplesetcoupled})$ we make the interesting observation that at the point where the derivative of $n_0(x)$ vanishes, $n_1(x)$ is not forced to flow to zero, but may in fact attain a non-trivial fixed point value.  This is demonstrated in the inset of figure \ref{gr1} (a) which shows $\log(n_1(x))$ as a function of $x$. One clearly sees a sharp peak at the point where the derivative of $n_0(x)$ vanishes approximately.  This peak only occurs for $\beta>1$ and moves to the right as $\beta$ is increased.  This is also consistent with the known result \cite{Wegner} that an off-diagonal element $m_{ij}$ of $H(\ell)$ decays roughly as $\exp(-(E_{i-1}-E_{j-1})^2 \ell)$. This suggests a connection between quantum phase transitions, the corresponding disappearance of an energy scale (gap)\cite{Sachdev} in the thermodynamic limit and the absence of decoupling in the Hamiltonian, also in the thermodynamic limit.
   
Upon further analysis of the flow equation one observes that the point where the derivative of $n_0(x)$ vanishes, must necessarily be a point of inflection.  The reason for this is that the derivative of $n_0(x)$ cannot change its sign as the flow equation for $n_1(x)$ will become unstable upon which $n_1(x)$ will grow exponentially, something that cannot happen as explained in section \ref{Intro} (see equation \ref{diagproof}). This explains why we only observe a plateau in figure \ref{gr1} (a) and no more drastic behaviour.  Note, however, from figure \ref{orderex} (a) that, as can happen at a point of inflection, the second order derivative of $n_0(x)$ does not exist, but changes sign. The fact that the dervative of $n_0(x)$ cannot change sign also explains the remark often made that the ordering of the eigenvalues is the same in the interacting and non-interacting systems.

Against the background of the discussion above, the presence of a point of inflection at higher energies for values of $\beta>1$ presents the interesting interpretation of a 'quantum phase transition' at higher energies. To illustrate that this is indeed a sensible interpretation we show the expectation value $1+\left<J_z\right>/j$ at $\beta=0.5, 1.5, 2.5$ as a function of $x$, i.e., as evaluated in the excited states, in figure \ref{orderex} (b).  One clearly sees a change in the behaviour at the point of inflection.  Indeed, after this point the behaviour of this quantity is linear, as is to be expected in the first phase, so that one can think of this point as a transition from the second phase, present at lower energies, to the first phase at still higher energies.  This interpretation is summarized in a 'phase diagram' in figure \ref{orderex} (c) which shows the $\beta$ dependancy of a subset of the negative, even eigenvalues. For $\beta\leq1$ the eigenvalues are confined between $0$ and $-j$. As $\beta$ increases first the groundstate (shown in bold) and then the excited states begin to cross the $E=-j$ phase boundary until eventually, in the large coupling limit, only the $\ket{E_j=0}$ state retains its first phase character. For finite $j$ successive eigenvalues show avoided level crossings on the phase boundary.  In the thermodynamic limit one would, however, expect that successive eigenvalues will coalesce as they cross the phase boundary; signalling a vanishing gap. Keeping in mind the symmetry $E_{n}=-E_{2j-n}$, these remarks can easily be adapted for the positive half of the spectrum as well.

\subsection{The order parameter}
\label{numorder}
Next we present results for the expectation values, computed according to the method described in section \ref{expval}. Figure \ref{gr1} (c) shows the order parameter $\Omega=1+\ave{H_0}/j$ as a function of $\beta$ together with the exact values for $j=1000$. Again we find excellent agreement, even at the phase transition, with absolute errors in the range of $5\times10^{-3}$. The second moment of $J_z$ is found by choosing $P=J_z^2/j^2$, for which $g(\ave{H_0},\ave{H_1},\ell=0)=\ave{H_0}^2/j^2$. Proceeding as before we obtain the expectation value of $P$ as a function of $\beta$, which appears, together with the exact result for $j=1000$, in figure \ref{gr1} (d). Note that the computation of the expectation values of different operators entails solving the {\it same} partial differential equation, but with different initial conditions.  In this sense the flow equation only knows about the operator being computed through the initial conditions.

\subsection{The structure of the eigenstates}
\label{eigenstatenumeric}

\begin{figure}[ht]
\begin{center}
	\epsfig{file=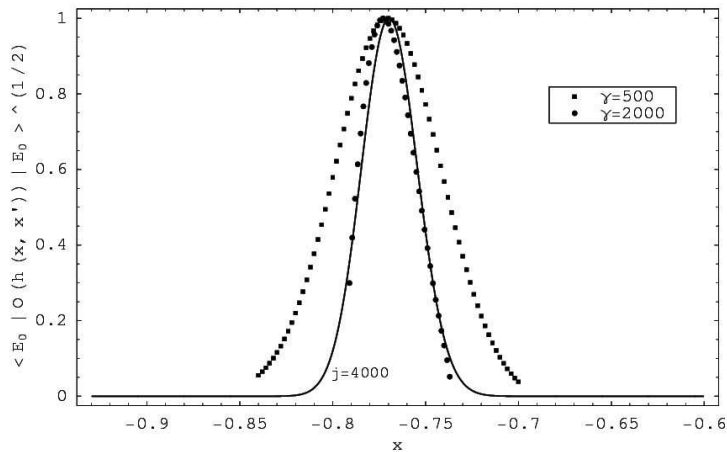,height=6cm,clip=,angle=0}\\
	\caption[eigstruc]{The expectation values of $O(h(x,x'))$ at $\beta=1.3$ for different $x'$. All results have been normalized to have the same maximum value.}
	\label{eigstruc}
\end{center}
\end{figure}
We follow the strategy discussed in section \ref{eigenstates} and consider the flow of the operator $O(h(x,\bar{x}))$ for $h(x,\bar{x})=\exp(-\gamma(\bar{x}-x)^2)$ and $\gamma=500,2000$. The resulting ground state expectation values for $\beta=1.3$ and a range of $\bar{x}$ values appear in figure \ref{eigstruc}. The exact values of $\left|\braket{xj}{E_0}\right|$, calculated from direct diagonalization, are shown as functions of $x$. Since the ground state lies within the subspace spanned by even basis states we completely neglect the coefficients corresponding to odd states. As before the flow equations cannot distinguish these two subspaces on this level, and only provide information on the average contribution of a range of both even and odd basis states. As expected, larger values of $\gamma$ provide a more accurate probing of the eigenstate, and for $\gamma=2000$ we obtain a good approximation for the form of the absolute wave function at large $j$ in the $J_z$ basis. 

In the first phase a very small subset of basis states has nonzero overlap with the ground state. This implies that, as a function of $x$, $\lim_{j\rightarrow\infty}\left|\braket{xj}{E_0}\right|$ will vary 
on a length scale which cannot easily be probed using this method. However, by considering $O(h(x)=x)=J_z/j$ and keeping in mind that $\inp{E_0}{J_z}{E_0}=-j\Leftrightarrow\ket{E_0}=\ket{-j}$ we conclude from figure \ref{gr1} (c) and (d) that $\ket{E_0}$ is approximately equal to $\ket{-j}$ when $\beta<1$.

\section{Summary and outlook}

We have shown that the flow equations based on continuous unitary transformations can be cast in the form of a general nonlinear partial differential equation involving one flow and two system specific auxiliary variables.  This approach is based on the observation that operators acting on finite dimensional Hilbert spaces, or bounded operators acting on infinite dimensional Hilbert spaces, may be written as polynomials in a set of irreducible operators, specifically two in the case of the flow equation.  Making an expansion to second order in the fluctuations yields the aforementioned flow equation. It is shown how the spectrum and expectation values can be extracted from the solution of this equation. The method is applied to the Lipkin model where all the calculations can be done easily and explicitly. 
Excellent results for the full spectrum and expectation values are obtained. It was illustrated how properties of the eigenstates can be investigated by considering the flow of certain diagonal operators. Further information about the eigenstates, for example the relative phases of expansion coefficients, could possibly be obtained by extending this strategy to off-diagonal operators, or by considering the flow of $U(\ell)$ itself. The flow equation also provides some interesting new insights into the nature of the quantum phase transition.   

A further possible extension of this approach is to study time dependent problems.  Of course, once the eigenvalues and eigenfunctions have been obtained to a reliable accuracy one can, in principle, determine the time dependency of the system.  We would, however, like to find a more economical way of computing the time dependency.  

Further possible applications to lattice models are currently being investigated.  In particular one-dimensional lattice models should be amenable to this approach as they are in structure very similar to the Lipkin model.  Indeed, the Lipkin model can be viewed as a lattice model with position dependent hopping amplitude and a linear potential.  

\ack

The authors acknowledge financial support from the National Research Foundation of South Africa.  We also acknowledge useful discussions with H B Geyer and W D Heiss.
    
\appendix
\section{Representations by irreducible sets}
\label{appa}
In section \ref{genrepsec} we made use of the fact that any flowing operator may be expressed in terms of operators coming from an irreducible set which contains the identity. Now we show why this is the case, using a result by von Neumann. First we establish some notation. Let ${\cal H}$ be a finite dimensional Hilbert space and denote by ${\cal B}({\cal H})$ the set of all matrix operators acting on it. The commutant ${\cal M}'$ of a set ${\cal M}\subseteq{\cal B}({\cal H})$ is defined as the set of operators which commute with all the elements of ${\cal M}$. The double commutant of ${\cal M}$ is defined by ${\cal M}''\equiv\left\{{\cal M}'\right\}'$. The following theorem \cite{Conway} provides the key:

\textbf{The Double Commutant Theorem:} \textit{If ${\cal A}$ is a subalgebra of ${\cal B}({\cal H})$ which contains the identity and is closed under hermitian conjugation then ${\cal A}={\cal A}''$.}

Now suppose ${\cal S}\subseteq{\cal B}({\cal H})$ is an irreducible set of hermitian operators, one of which is the identity, and ${\cal A}$ the subalgebra of ${\cal B}({\cal H})$ spanned by all the products of operators from ${\cal S}$. By Schur's lemma ${\cal S}'={\cal A}'=\left\{\lambda I\,|\,\lambda\in\mathbb{C}\right\}$ and thus ${\cal A}''={\cal B}({\cal H})$. It follows from the theorem above that ${\cal A}={\cal B}({\cal H})$, i.e. any operator acting on ${\cal B}({\cal H})$ may be written in terms of operators coming from ${\cal S}$.
\section{Calculating expectation values with respect to coherent states}
\label{appb}
Define $|z\rangle =(1+z z^\ast)^{-j}|z)=N^{-j}|z)$ where $|z)=\exp(zJ_+)\ket{-j}$ is the unnormalized coherent state. The differential operator representations for the ${\rm su}(2)$ generators are given by \cite{Perelomov}:
\begin{eqnarray}
J_z|z)=\left(-j+z\frac{\partial}{\partial z}\right)|z)=\hat{J}_z\left(z,\partial_z\right)|z)\\
J_+|z)=\frac{\partial}{\partial z}|z)=\hat{J}_+\left(z,\partial_z\right)|z)\\
J_-|z)=\left(2zj-z^2\frac{\partial}{\partial z}\right)|z)=\hat{J}_-\left(z,\partial_z\right)|z).
\end{eqnarray}
We will always consider $z$ and $z^*$ to be distinct variables, and that $(z|$ is a function of $z^*$ only. When calculating expectation values of the double commutators in section \ref{genrepsec}, one encounters expressions of the form 
\begin{equation}
\langle z|H_{n_1}H_{n_2}...H_{n_\alpha}|z\rangle=\frac{(z|H_{n_1}H_{n_2}\ldots H_{n_\alpha}|z)}{(z|z)}
\label{genexp}
\end{equation}
where $n_i=0  \ {\rm or} \ 1$. These can easily be computed by replacing each operator by its differential representation to obtain
\begin{equation}
	\langle z|H_{n_1}H_{n_2}...H_{n_\alpha}|z\rangle=\frac{1}{(z|z)}\hat{H}_{n_\alpha}\left(z,\partial_z\right)\ldots \hat{H}_{n_1}\left(z,\partial_z\right)(z|z),
\end{equation}
where the orders of the operators have been reversed. As a result these expectation values can be expressed as some rational function of $z=r\exp(i\theta)$ and $z^*=r\exp(-i\theta)$. Applying this method to $\ave{H_0}$ and $\ave{H_1}$ we obtain
\begin{eqnarray}
\langle z|H_0|z \rangle =j\left(\frac{r^2-1}{r^2 +1}\right)\ {\rm and} \ \ \langle z|H_1|z \rangle =\frac{(2j-1)r^2\cos(2\theta)}{(1+r^2)^2}.
\label{hifuncrt}
\end{eqnarray}
Using these equations to express $r$ and $\theta$ in terms of $\ave{H_i}$ we may consider expectation values of the form of equation (\ref{genexp}) as functions of these averages.
For example
\begin{eqnarray}
\fl\langle z|[[\Delta H_0,\Delta H_1],(\Delta H_0)^2]|z\rangle=8jxy,
\label{cancelexample}
\end{eqnarray}
where $\langle H_0 \rangle=jx$ and $\langle H_1 \rangle =jy$.

\section{Scaling behaviour of fluctuations}
\label{appc}
We wish to show that 
\begin{equation}\label{aa}
\langle z|\Delta H_{n_1}\Delta H_{n_2}...\Delta H_{n_t}|z\rangle \sim j^{\lfloor \frac{t}{2} \rfloor}
 \ \ {\rm for}  \ \forall  \  t\in \mathbb{N}  \ {\rm where}  \ n_i\in \left\{0,1\right\}.
\end{equation}
From direct calculation this is found to hold for $t=2$, and we employ induction to obtain the general result. Assuming that this holds for all products of $k\leq t$ fluctuations, the induction step consists of adding either $\Delta H_0$ or $\Delta H_1$ to a general product $M=\Delta H_{n_1}\Delta H_{n_2}...\Delta H_{n_t}$ and then proving the result for $\inp{z}{M\Delta H_i}{z}$.\\

First we add an extra $\Delta H_0$. Using the results from the previous section we may write
\begin{eqnarray}
\fl (z|M\Delta H_0|z)&=&-j(z|M|z)+(z|M z\partial_z|z)-\langle H_0 \rangle(z|M|z)\\
\fl &=&-2j z z^\ast (1+z z^\ast)^{-1}(z|M|z)-z(z|(\partial_z M)|z)+z\partial_z(z|M|z)
\end{eqnarray}
where (\ref{hifuncrt}) and $(z|M z\partial_z|z)=z(\partial_z(z|M|z)-(z|(\partial_z M)|z))$ were used. It should be remembered that $M$ contains a $z$ dependency through the average appearing in each $\Delta{H_i}$. Dividing by $(z|z)$ leads to
\begin{eqnarray}\label{cc}
\fl \langle z|M\Delta H_0|z\rangle=-2j z z^\ast (1+z z^\ast)^{-1}\langle z|M|z\rangle-z\langle z|(\partial_z M)|z\rangle \nonumber \\
+\frac{z}{(1+z z^\ast)^{2j}}\partial_z(z|M|z).
\end{eqnarray}
The last term can be rewritten using
\begin{equation}
\frac{1}{(1+z z^\ast)^{2j}}\partial_z(z|M|z)=\partial_z\langle z|M|z\rangle +\frac{2j z^\ast}{(1+z z^\ast)}\langle z|M|z\rangle,
\end{equation}
which is just the product rule, to obtain
\begin{equation}\label{dd}
\langle z|M\Delta H_0|z\rangle=z \partial_z\langle z|M|z\rangle-z\langle z|(\partial_z M)|z\rangle.
\label{finform}
\end{equation}
Note the important cancellation of terms proportional to $j\langle z|M|z\rangle$. From the induction hypothesis, and the fact that $\langle z|M|z\rangle$ is polynomial in $j$, the first term in (\ref{finform}) will be of order $j^{\lfloor \frac{t}{2} \rfloor}$. The second term becomes
\begin{equation}\label{ee}
z\sum_{i=1}^t \langle z|\Delta H_{n_1}...\Delta H_{n_{i-1}}\Delta H_{n_{i+1}}...\Delta H_{n_t}|z\rangle (\partial_z\langle H_{n_i}\rangle).
\label{sumterm}
\end{equation}
Taking into account the expressions for $\langle H_0 \rangle$ and $\langle H_1 \rangle$, we conclude that each term in equation (\ref{ee}) will have order $j^{1+\lfloor \frac{t-1}{2} \rfloor}$, which reduces to $j^{\lfloor \frac{t+1}{2} \rfloor}$ in both the cases where $t$ is odd and even. Thus, for any product $M$ of $t$ fluctuations we arrive at
\begin{equation}\label{ff}
\langle z|M\Delta H_0|z\rangle\sim j^{\lfloor \frac{t+1}{2} \rfloor},
\end{equation}
which concludes the induction step.

Exactly the same procedure is followed when adding a $\Delta H_1$, although more algebra is required as $\hat{H_1}(z,\partial_z)$ now contains second order derivatives to $z$. The final result remains unchanged:
\begin{equation}
\langle z|m\Delta H_1|z\rangle\sim j^{\lfloor \frac{t+1}{2} \rfloor}.
\end{equation}

In this case these results make exact the general notion that relative fluctuations scale like powers of one over the system size. Finally we mention that when calculating the expectation values of a double commutator there is often a cancellation of leading order terms. This can be seen in equation (\ref{cancelexample}), where the sum of terms of order $j^2$ turns out to be of order $j$.

\end{document}